\documentclass[aps,pra,twocolumn,showpacs,superscriptaddress]{revtex4-1}
\usepackage[]{graphicx}
\usepackage{bm}
\usepackage{amsmath}

\usepackage{color}

\usepackage[colorlinks=true,linkcolor=blue,citecolor=blue,urlcolor=blue]{hyperref}

\providecommand{\bra}[1]{\langle #1 \rvert}
\providecommand{\ket}[1]{\lvert #1 \rangle}

\providecommand{\ketbra}[2]{\lvert  #1\rangle \langle #2 \rvert}
\providecommand{\beq}{\begin{equation}}
\providecommand{\eeq}{\end{equation}}
\providecommand{\bea}{\begin{eqnarray}}
\providecommand{\eea}{\end{eqnarray}}
\begin{document}

\title{Coupling a single Nitrogen-Vacancy center to a superconducting flux qubit in the far off resonance regime}

\author{Tom Douce}
\affiliation{Laboratoire Mat\'eriaux et Ph\'enom\`enes Quantiques, Universit\'e Paris Diderot, CNRS UMR 7162, 75013, Paris, France}
\author{Michael Stern}
\affiliation{Quantronics group, SPEC, IRAMIS, DSM, CEA Saclay, 91191 Gif-sur-Yvette, France}
\affiliation{Quantum Nanoelectronics Laboratory, MS2 Tech., BINA, Bar-Ilan University, Ramat Gan, 52900, Israel}
\author{Nicim Zagury}
\affiliation{Instituto de F\'{\i}sica, Universidade Federal do Rio de Janeiro. Caixa Postal 68528, 21941-972 Rio de Janeiro, RJ, Brazil}
\author{Patrice Bertet}
\affiliation{Quantronics group, SPEC, IRAMIS, DSM, CEA Saclay, 91191 Gif-sur-Yvette, France}
\author{P\'erola  Milman}
\affiliation{Laboratoire Mat\'eriaux et Ph\'enom\`enes Quantiques, Universit\'e Paris Diderot, CNRS UMR 7162, 75013, Paris, France}

 
\begin{abstract}
We present a theoretical proposal to couple a single Nitrogen-Vacancy (NV) center to a superconducting flux qubit (FQ) in the regime where both systems are off resonance. The coupling between both quantum devices is achieved through the strong driving of the flux qubit by a classical microwave field that creates dressed states with an experimentally controlled characteristic frequency. We discuss several applications such as controlling the NV center's state by manipulation of the flux qubit, performing the NV center full tomography and using the NV center as a quantum memory. The effect of decoherence and its consequences to the proposed applications are also analyzed. Our results provide a theoretical framework describing a promising hybrid system for quantum information processing, which combines the advantages of fast manipulation and long coherence times. 
\end{abstract}
\pacs{03.67.Bg, 42.50.Ex}
\vskip2pc 
 
\maketitle
\section{Introduction}

Electronic spins in semiconductors such as NV centers in diamond or
impurities in silicon are characterized by their low decoherence rates
$\Gamma_{s}$ \cite{Bar-Gill,muhonen,Saeedi,wolfowicz}. They would be excellent
candidates for realizing an operating quantum processor if only there
was a way to reliably entangle distant spins~\cite{bernien13,awschalom07}. Recently, it has been
proposed to solve this issue by using a superconducting circuit as
a quantum bus \cite{Marcos,Twamley}. This approach would require to
reach the strong coupling regime where the coupling strength $g$
between the spin and the circuit is larger than their respective decoherence
rates $\left(g\gg\left\{ \Gamma_{s},\Gamma_{circuit}\right\} \right)$.
To achieve this goal, the coupling constant $g$ should be increased
by several orders of magnitude compared to the current state of the
art where $g$ is typically of the order of a few Hz \cite{Kubo,Zhu}. 

\begin{figure}[h]
\includegraphics[scale=0.5]{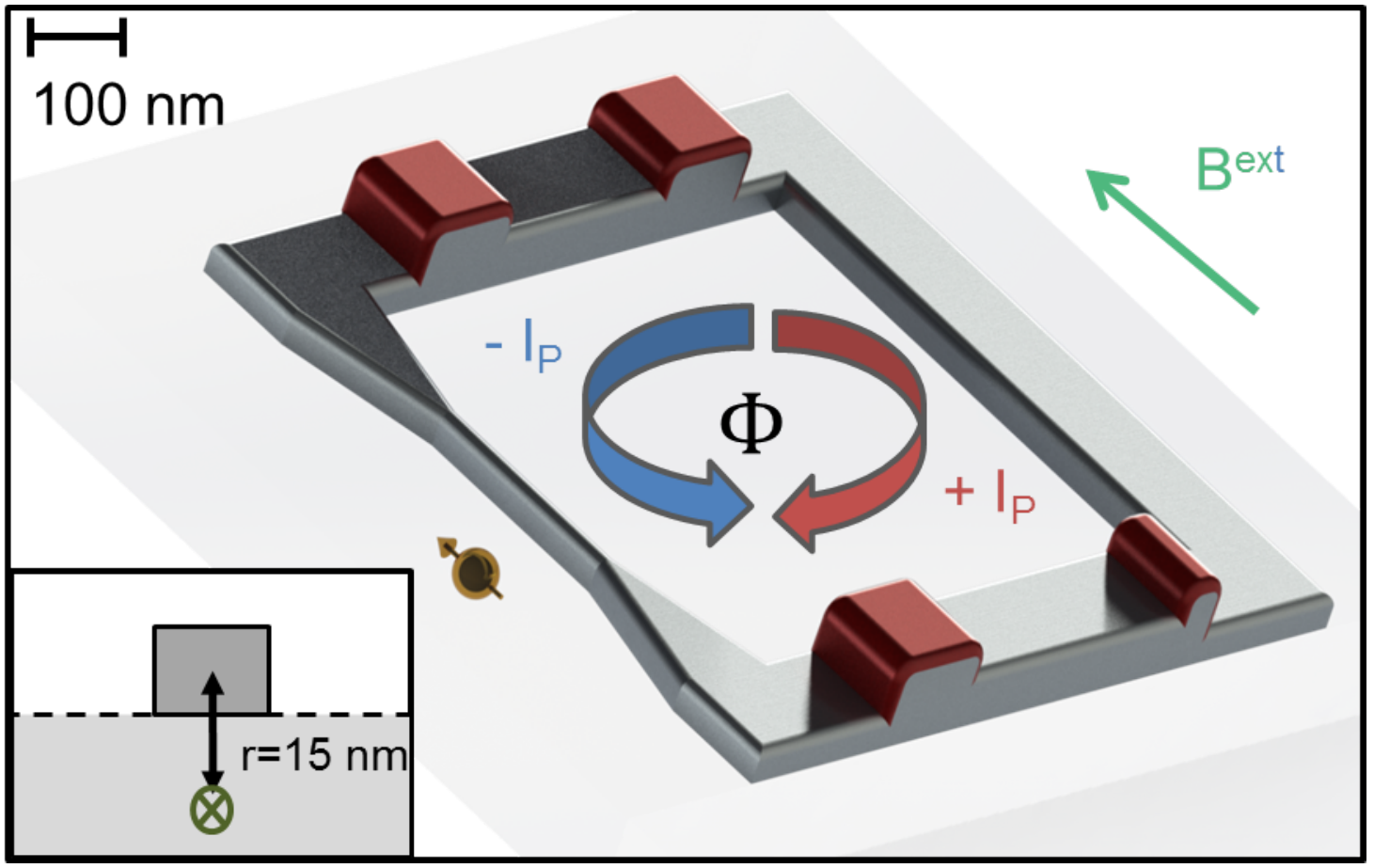}
\caption{\label{spinFQ} (Color online)  Schematic view of a flux qubit coupled to a single spin. The flux qubit consists of a superconducting loop with a constriction and intersected by four Josephson junctions (in red). Three of the junctions are identical while the fourth junction is smaller than others. When the flux threading the loop $\Phi$ is close to half a flux quantum, the system behaves as a two level system The Nitrogen Vacancy center represented as a golden arrow is situated in the close vicinity ($r=15$ nm) of the constriction to maximize the coupling between the two systems. The application of an external magnetic field $B_{\rm ext}$ in the direction of the NV axis lifts the degeneracy between the $m_S=\pm1$ states.}
\end{figure}

The coupling constant $g$ can be greatly increased by using flux
qubits instead of a linear superconducting resonator. Indeed, flux
qubits~\cite{mooji99,vanderWal00,bertet05,forndiaz10,Orlando99,PREP,Nakamura06,Bylander11} are characterized by a macroscopic permanent current $I_{P}$
typically of the order of several hundreds of nA. This current flows
in the loop of the qubit and generates therefore a large magnetic
dipole which allows an efficient coupling to spins. Bringing the spin
at a distance of $\sim20$ nm from the flux qubit can be achieved
by fabricating an ultra-narrow superconducting wire and aligning it
very precisely with the spin (see Fig.~\ref{spinFQ}). This would allow increasing the coupling
by several orders of magnitude and reaching a coupling constant $g/2\pi\sim100$~kHz, a value much larger than recently reported spin decoherence
rates \cite{Bar-Gill,muhonen}. However, flux qubits suffer from
a severe limitation. Their large magnetic dipole makes them very sensitive
to flux noise which limits their coherence times. To cope with this problem, one should tune the flux
threading the loop $\Phi$ precisely at half a flux quantum ($\Phi=\Phi_{0}/2$),
the so-called optimal point of the flux qubit~\cite{bertet05}. At this point its energy does not depend on the flux at first order and is equal to the gap $\hbar\Delta$. 

The coupling of a spin -- whose transition energy $\hbar\omega_{s}$
is given by Nature -- with a flux qubit is optimal for $\omega_{s}=\Delta$.
Unfortunately, reaching this target is a challenging task.
The value of $\Delta$ is an exponential function~\cite{Orlando99} of the parameters
of the junctions which form the qubit and it is therefore poorly controlled. It is possible to design a tunable flux qubit by replacing one of the junctions by a SQUID~\cite{Paauw}. However, coherence times required for single spin coupling have not been reported so far in this geometry. Applying an external magnetic field to tune the spin close to $\Delta$
frequency has also many limitations. The critical field of Aluminium
is $\sim100$ G and therefore limits the tunability of the spin frequency should be limited to 100-200 MHz. As a consequence the two resonance frequencies $\omega_s$ and $\Delta$ may
be highly detuned. Typically, one can expect that $(\Delta-\omega_{s})/2\pi\sim300-500$
MHz, prohibiting the use of the direct resonant interaction to entangle the two systems. In this work, we propose a theoretical strategy to cope with
this issue and couple efficiently both systems even with such a  detuning.

\section{The model}

\subsection{Effective Hamiltonian}\label{sec1}

We now present our model, starting with NV centers, the FQ and the coupling mechanism. NV centers are diamond color centers consisting of a Nitrogen substitutional impurity next to a vacancy~\cite{Gruber,Manson}. In their negatively-charged state, they have an electronic ground state with a spin $S=1$ which is described by the following Hamiltonian (neglecting the effect of strain~\cite{Neumann}):
\beq
\hat H^{s}= \hbar D S_z^2+\hbar\gamma_e \vec{S}\cdot\vec{B}^{\rm ext},
\eeq
where $\vec S$ are the dimensionless spin-1 operators, $D/(2\pi)\approx 2.88$ GHz is the zero-field splitting ground state and the rightmost term is the Zeeman interaction of the electronic spin of gyromagnetic ratio $\vert\gamma_e\vert/2\pi = 28$ GHz/T. The quantization $z$-axis is set in the Nitrogen Vacancy axis. $B^{\rm ext}$ is an external magnetic field applied to lift the degeneracy of the $\{m_S=\pm1\}$ levels. For an antiparallel field, the NV center is well described by a two-level system Hamiltonian in the $\{m_S=0,m_S=+1\}$ basis:
\beq  
\hat H^{s}=\hbar\omega_s \hat\sigma_z^{s}/2\, ,
\eeq
where $\omega_{{s}}=D+\gamma_eB^{\rm ext}$,  $\hat\sigma_z^{s}=\ketbra{1}{1}-\ketbra{0}{0} $ with $\ket 0$ and $\ket 1$ corresponding respectively to $m_S=0$ (lower state) and $m_S=1$.
  
The flux qubit Hamiltonian can also be described by a two level system \cite{Orlando99}:
\beq
\hat H^{fq}=\hbar \epsilon \hat \sigma_x^{fq}/2+\hbar\Delta\hat \sigma_z^{fq}/2\, ,
\eeq
where $\epsilon=2I_p(\Phi-\Phi_0/2)$. The NV is in a position such that 
 the field from the flux qubit is normal to the quantization axis of NV spin $\vec S .$ 
   Therefore the Hamiltonian for the two interacting systems can be written as  
 \beq
  \hat H= \hbar\omega_s \hat \sigma_z^{s}/2 +\hbar\epsilon\sigma_x^{fq}/2 + \hbar\Delta\sigma_z^{fq}/2 +\hbar g\sigma^{s}_x\sigma_x^{fq},
 \eeq
where the fourth term is the Zeeman interaction of the electronic spin due to the magnetic field generated by the flux qubit. The coupling constant $g$ is given by $g\simeq\gamma_e\mu_0I_p/(\sqrt2\cdot2\pi r)$ where $\mu_0$ is the vacuum permeability and $r$ the distance between the spin and the constriction shown in Fig.~\ref{spinFQ}. By setting $r=15$ nm and $I_p=500$ nA, we get $g/2\pi\simeq100$ kHz. In the following, we will assume that $\Phi = \Phi_0 / 2$, in order to benefit from the longest flux-qubit coherence times.

The flux qubit is subjected to an additional microwave magnetic field of frequency $\omega$ in a direction normal to the plane of the persistent currents. The NV spin state has a much smaller magnetic dipole and therefore we will assume that it is not driven  directly by the microwave field. Thus the Hamiltonian reads:
 \bea\label{Hs}
  \hat H&=& \hbar\omega_s \hat \sigma_z^{s}/2 +\hbar\Delta \hat\sigma_z^{fq}/2  +\hbar g \hat \sigma^{s}_x 
  \hat\sigma_x^{fq}\nonumber \\ 
  &+&\hbar \Omega \hat \sigma_x^{fq}  \cos(\omega t)
\eea
where $\Omega=I_PAB_0$ is the Rabi frequency of the flux qubit of area $A$ and persistent current $I_P$ driven by the classical microwave field $B(t)=B_0\cos(\omega t)$.

\begin{figure}
\includegraphics[scale=.44]{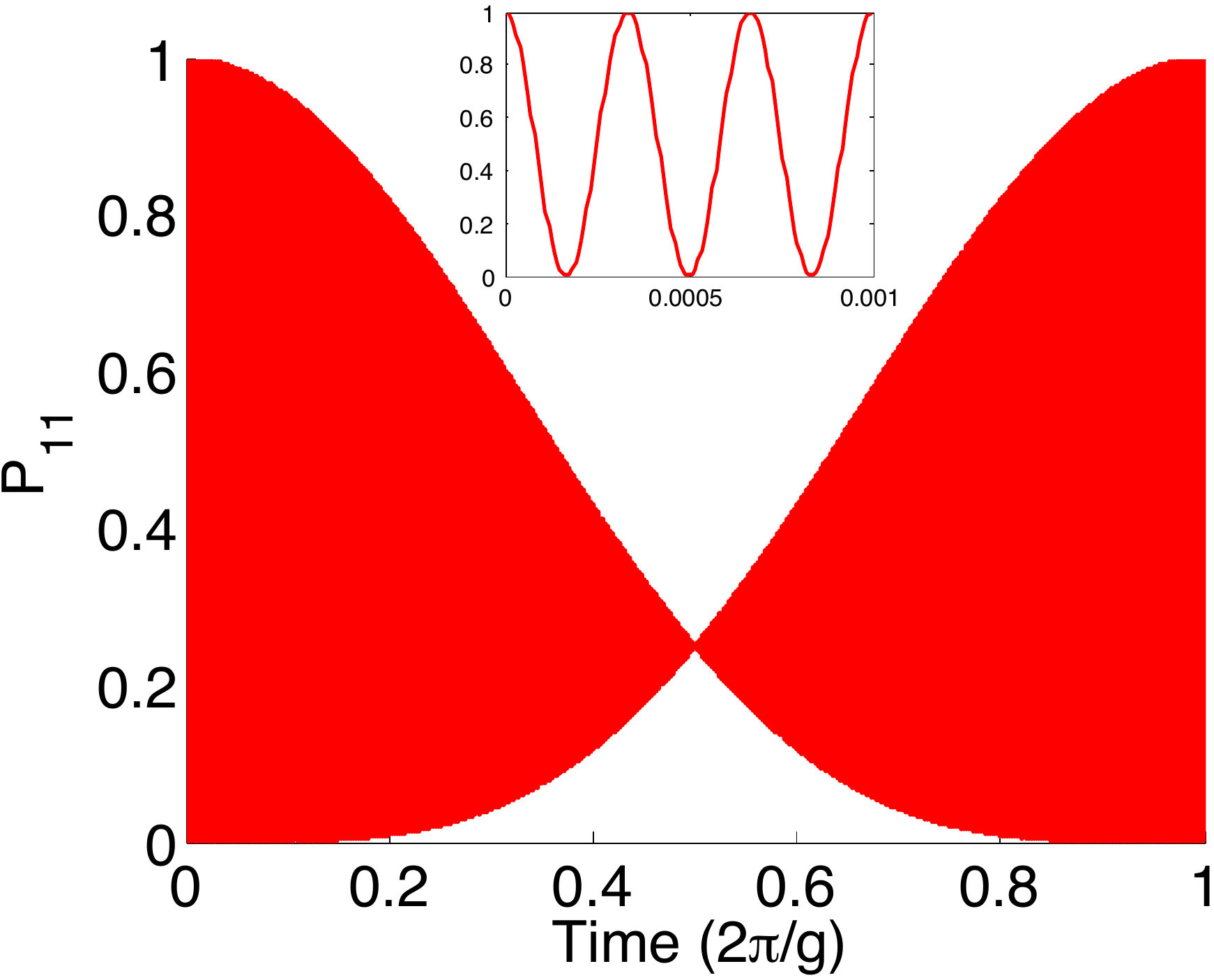}
\caption{\label{figRabi}(Color online) Probability of finding the composite system in state $\ket1\ket1$ as a function of time (in units of $2\pi/g$). We numerically solve the Schr\"odinger equation with Hamiltonians $\hat H_0$ and $\hat H_{\rm int}$ from Eq.~\ref{lab}. We can see the slow dynamics induced by the magnetic coupling while the inset displays the fast Rabi oscillations occurring with a much shorter time scale. The set of parameters is $\omega_{s}=28800g$, $\Delta=25800g$ and $\Omega=-\delta$.}
\end{figure}

From (\ref{Hs}) we can move to  a frame rotating with the classical field's frequency $\omega$, which is set to be the same as the flux qubit transition frequency -- namely $\omega=\Delta$. This is accomplished  through the  unitary operator transformation $U_s(t)={\rm e}^{-i\Delta t (\hat \sigma_z^{s}+\sigma_z^{fq} )/2}\, U_r(t),  $ connecting  the time evolution operators in the Schroedinger and rotating frame pictures. We define  $\delta=\omega_{s}-\omega=\omega_{s}-\Delta$ and the raising/lowering operators in the $\sigma_z$ basis, $\hat \sigma_{\pm}=\left (\hat \sigma_x \pm i\hat \sigma_y\right )/2$. By setting $\Omega \simeq |\delta|$, we identify two terms in the resulting Hamiltonian in the rotating frame: 
\begin{equation}
\hat H_r=\hat H_0+\hat H_{\rm int},
\end{equation}
with
\begin{eqnarray}\label{lab}
&&\hat H_0=\hbar\frac{\delta}{2}  \hat \sigma_z^s+\hbar \frac{\Omega}{2} \hat \sigma_x^{fq} \\
&&\hat H_{\rm int}=\hbar g (\hat \sigma_+^s{\rm e}^{i\Delta t}+\hat \sigma_-^s{\rm e}^{-i\Delta t})(\hat \sigma_+^{fq}{\rm e}^{i\Delta t}+\hat \sigma_-^{fq}{\rm e}^{-i\Delta t}).\nonumber
\end{eqnarray} 
We can thus consider that the classical drive ``dresses" the flux qubit \cite{KIKE, NOS}. From now on this representation will be referred to as the laboratory frame. In order to understand the physical meaning of this time-dependent interaction Hamiltonian, we plotted in Fig.~\ref{figRabi} the time evolution of the probability of finding both systems in their excited states $\ket1\ket1$. The figure displays two distinct time scales: the first one is given by the Rabi oscillations of frequency $\Omega$ induced by the classical field. The second one comes from the magnetic coupling between the flux qubit and the spin. For the numerical simulation we have set $\omega_{s}=28800g$ and $\Delta=25800g$ so that $\delta=3000g$, with $\Omega=-\delta$.

We now move to the interaction picture with respect to $\hat H_0$ in order to get an approximate expression for the interaction Hamiltonian. We have:
\begin{eqnarray}\label{inter}
&&\hat H_I=\hbar \frac{g}{2}\left \{\hat \sigma_+^s  {\rm e}^{i(2\Delta+\delta )t}  \left [\hat \sigma_x^{fq}-{\rm e}^{i\Omega t}\left (\hat \sigma_z^{fq}-i\hat \sigma_y ^{fq}\right )/2 \right. \right. \nonumber \\
&&\left. \left.+{\rm e}^{-i\Omega t} \left (\hat \sigma_z^{fq}+i\hat \sigma_y^{fq} \right )/2 \right ]+\hat \sigma_+^s  {\rm e}^{i\delta t} \left [ \hat \sigma_x^{fq}- \right. \right. \\
&&\left. \left. {\rm e}^{-i\Omega t}\left (\hat \sigma_z^{fq}+i\hat \sigma_y^{fq} \right )/2 + {\rm e}^{i\Omega t} \left (\hat \sigma_z^{fq}-i\hat \sigma_y^{fq} \right )/2 \right ]+{\rm H.c}\right \}. \nonumber 
\end{eqnarray}
Let us define $\hat \sigma_{+,x}=\ket{+}\bra{-}=(\hat \sigma_z-i\hat \sigma_y)/2$ and  $\hat \sigma_{-,x}=\ket{-}\bra{+}=(\hat \sigma_z+i\hat \sigma_y)/2$, where $\ket{\pm}=(\ket1\pm\ket0)/\sqrt2$. We recall that we focus on the case with $\Omega\simeq |\delta|$ in Eq.~(\ref{inter}). Assuming also the conditions ($\Delta, \Omega, \delta \gg g$) we can perform the so-called rotating wave approximation and are led to the effective Hamiltonian:
\begin{equation}\label{Heff}
\hat H_{\rm eff}(t)=\pm\hbar\frac{g}{2}\left ( \hat \sigma_+^s \hat \sigma_{\pm,x}^{fq}{\rm e}^{i(\Omega\pm\delta)t}+\hat \sigma_-^s \hat \sigma_{\mp,x}^{fq}{\rm e}^{-i(\Omega\pm\delta)t} \right ),
\end{equation}
The effective Hamiltonian (\ref{Heff}) describes a closed two qubit system that can be diagonalized and  studied analytically. It can lead to a number of applications that will be further detailed in the next Sections. We stress that in the case of exact matching condition $\Omega=\pm\delta$ the effective Hamiltonian becomes time-independent. In the following we set $\Omega=-\delta$ so that $\hat H_{\rm eff}=\hbar\frac{g}{2}\left ( \hat \sigma_+^{s} \hat \sigma_{+,x}^{fq}+\hat \sigma_-^s \hat \sigma_{-,x}^{fq} \right )$. Adapting the results to $\Omega=\delta$ is straightforward.

Supposing that the spin-flux qubit bipartite system is initially in a separable  pure state, it can be written as:
\beq \label{init}
\ket{\psi(0)}=(\cos{\theta}\ket{0}+{\rm e}^{i\varphi}\sin{\theta}\ket{1})(\cos{\alpha}\ket{+}+{\rm e}^{i\phi}\sin{\alpha}\ket{-}). 
\eeq
In (\ref{init}), we dropped subscripts and used a convention that will be kept the same throughout this manuscript:   the spin's state is expressed  in the basis of $\hat \sigma_z^s$ eigenstates, $\ket{0}, \ket{1}$ and the flux qubit state  is expressed in the basis of $\hat \sigma_x^{fq}$ eigenstates, $\ket{+}, \ket{-}$. This basis choice is well adapted to the coupling induced by  (\ref{Heff}). Moreover, it provides a clear distinction between each party's states without using auxiliary labels. The parameters $\theta$ and $\varphi$ characterizing the spin's state are unknown and cannot be controlled or manipulated without the coupling (\ref{Heff}). However, the parameters  $\alpha$ and $\phi$ can be experimentally controlled and the flux qubit can be prepared in an arbitrary initial state by combining classical pulses with different phases, intensity and duration  \cite{PREP}.  The time evolution of the initial state (\ref{init}) subjected to Hamiltonian (\ref{Heff}) is: 
  \bea \label{time}
&&\ket{\psi(t)}=\cos{\theta}\cos{\alpha}\ket{0,+}+{\rm e}^{i(\phi+\varphi)}\sin{\theta}\sin\alpha\ket{1,-} \nonumber\\
&&+{\rm e}^{i\varphi}\sin\theta\cos\alpha(-i\sin\frac{gt}2\ket{0,-}+\cos\frac{gt}2\ket{1,+}) \nonumber\\
&&+{\rm e}^{i\phi}\cos\theta\sin\alpha(\cos\frac{gt}2\ket{0,-}-i\sin\frac{gt}2\ket{1,+}). 
\eea

\begin{figure}[h]
\includegraphics[scale=.23]{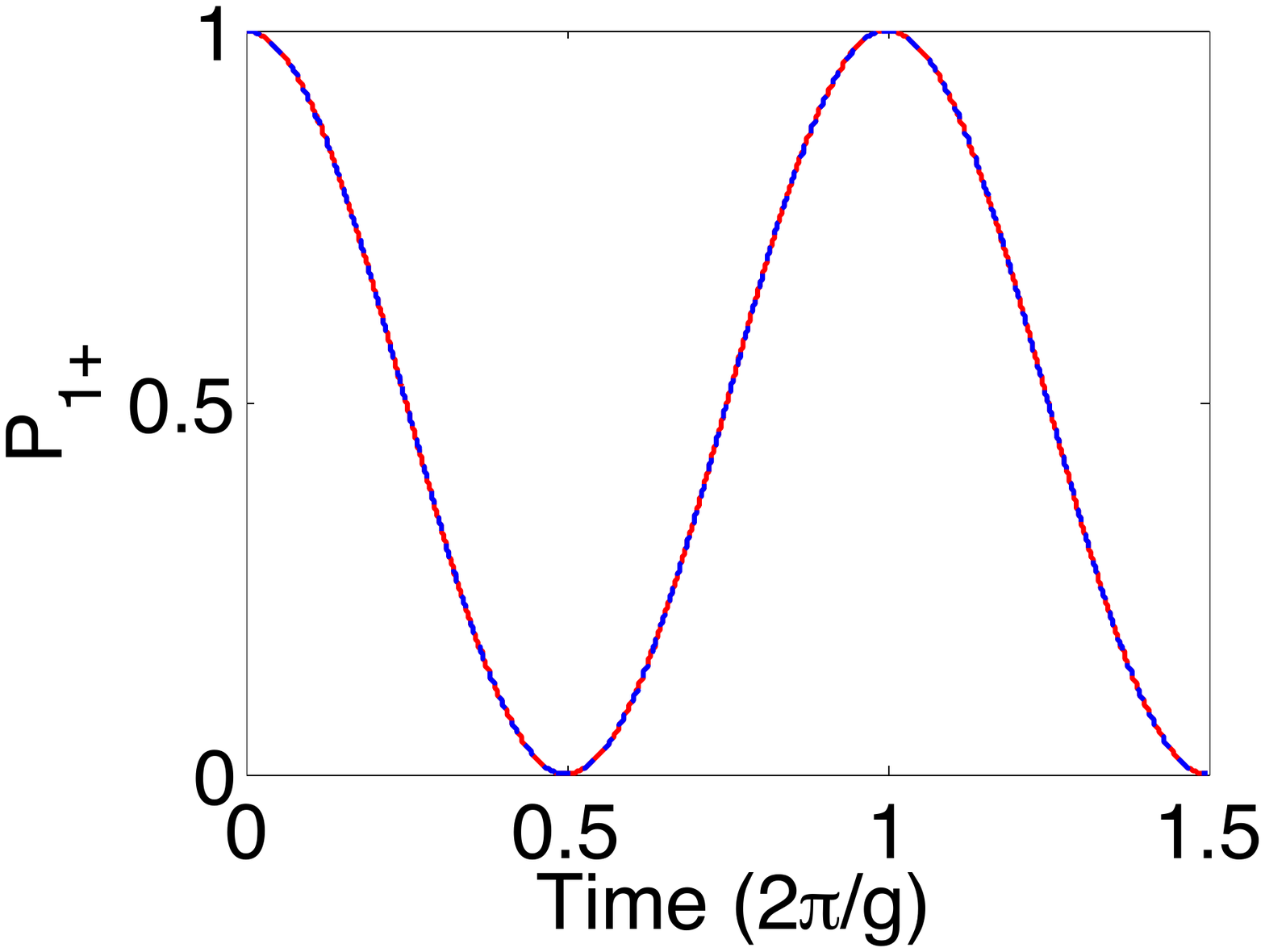}
\includegraphics[scale=.23]{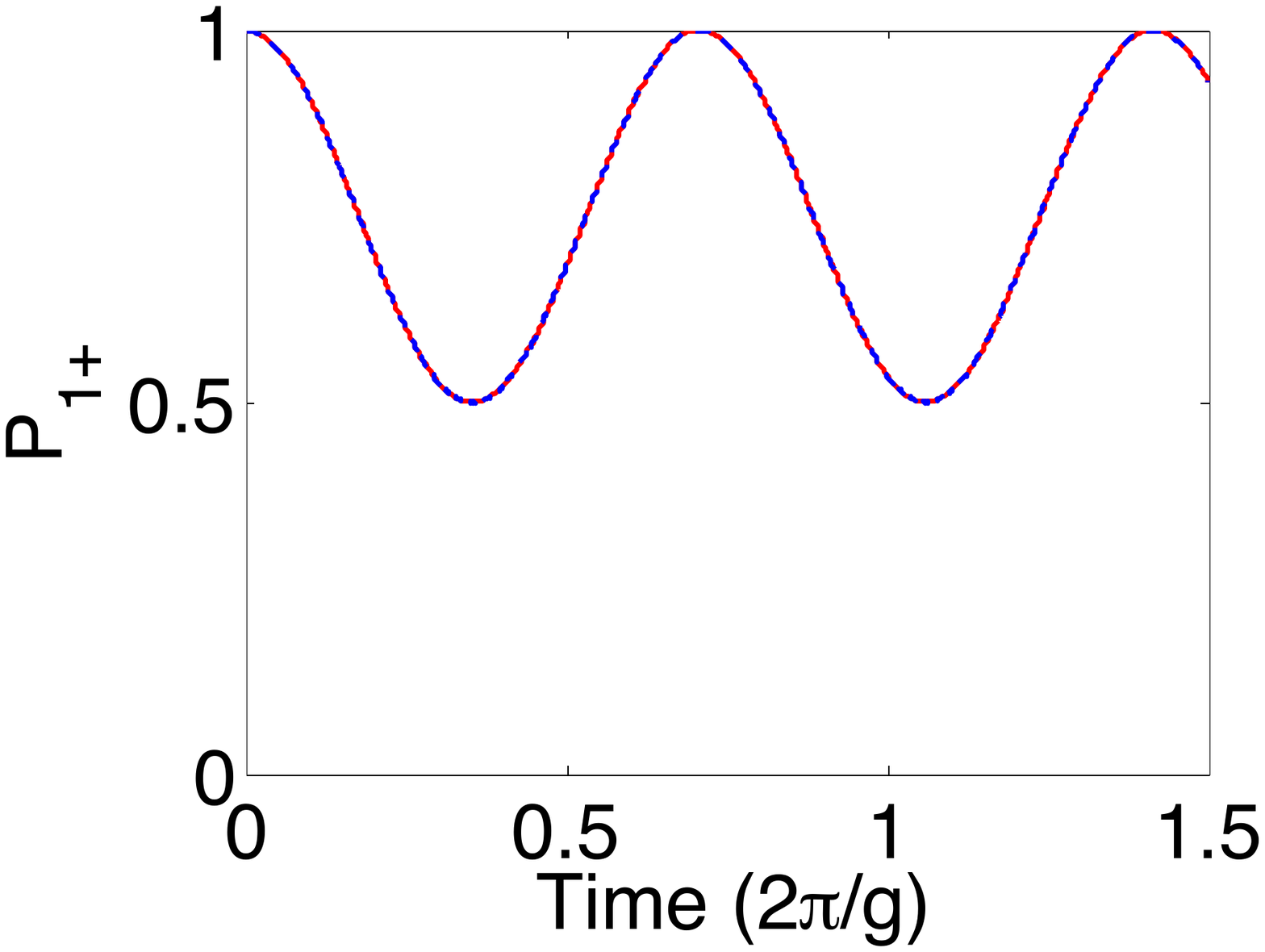}
\caption{(Color online) Probability of finding the composite system in state $\ket1\ket+$ as a function of time (in units of $2\pi/g$). In dashed blue we show the exact time evolution of population while in continuous red the one obtained using the effective Hamiltonian (\ref{Heff}). Left: Case of perfect resonance, $\Omega=-\delta$. Right: Non-resonant coupling $\delta+\Omega=g$. We can see that in both cases the effective evolution matches the exact one with very good accuracy. \label{fig1}} 
\end{figure}

In order to test the validity of our model, we compare the time evolution induced by the effective Hamiltonian (\ref{Heff}) to the evolution under the action of the exact Hamiltonian in the interaction picture (\ref{inter}). This can be done by comparing the numerical integration of the Schr\"odinguer equation using  (\ref{inter}) to the analytical diagonalization of (\ref{Heff}). We can study as an example the case where  $\theta=\pi/2$ and $\alpha=0$. In this case, the initial state is $\ket{1}\ket{+}$ using the basis convention. This state is an eigenstate of the free Hamiltonian $\hat H_0$, so the interaction Hamiltonian is solely responsible for the dynamics. The time evolution of the population in $\ket1\ket+$ under either $\hat H_{\rm eff}$ or $\hat H_{\rm int}$ is shown in Figure \ref{fig1}, with $\Omega=-\delta$ (left) and $\Omega=-\delta + g$ (right). We can see that indeed, both curves display a very good overlap. Thus the effective Hamiltonian is a good approximation to the coupling between both systems, even if the resonance is not perfect, {\it i.e.} $\Omega \neq \pm \delta$, in this case leading to a reduction of the fringes visibility.

The tuning of the resonance can be done by continuously changing the intensity and phase of the classical dressing field until coupling between the NV center and the flux qubit is optimal, as  in a ``quantum radio". This point  will be further discussed in the Section \ref{detecting}. 

We should also mention that similar ideas have been proposed to couple superconducting qubits with widely different frequencies~\cite{rigeti,ashhab}.

\subsection{Decoherence}\label{Decoherence}

The dynamics of the coupled spin-flux qubit system will be influenced by the presence of noise, that creates population losses and dephasing. In order to describe the proper dynamical equation for the density matrix of the system, we will start from a microscopic description of the flux qubit-environment interaction and then derive the corresponding Lindblad equation. Given the typical decay rates of a flux qubit compared to those of a NV center, we will neglect contributions to decoherence induced by the dephasing of the latter. Our microscopic model is thus made of two interactions: a dissipative process conveyed by the qubit $\hat \sigma_x$ operator and a dephasing one due to $\hat \sigma_z$. More precisely we have:
\begin{equation}
\hat H_{\mathrm{diss}}=\hbar\sum_k\gamma_k\hat \sigma_x^{fq}(\hat b_k+\hat b_k^\dagger)
\end{equation}
and~\cite{breuer}
\begin{equation}
\hat H_{\mathrm{deph}}=\hbar\sum_{k}\lambda_k\hat \sigma_z^{fq}(\hat c_k+\hat c_k^\dagger)
\end{equation}
where the $\hat b_k^{(\dagger)}$'s and $\hat c_k^{(\dagger)}$'s are bosonic annihilation (creation) operators of the modes of the environment. In our study we will assume the environment is in thermal equilibrium at zero temperature.

\begin{figure}[h]
\includegraphics[scale=0.42]{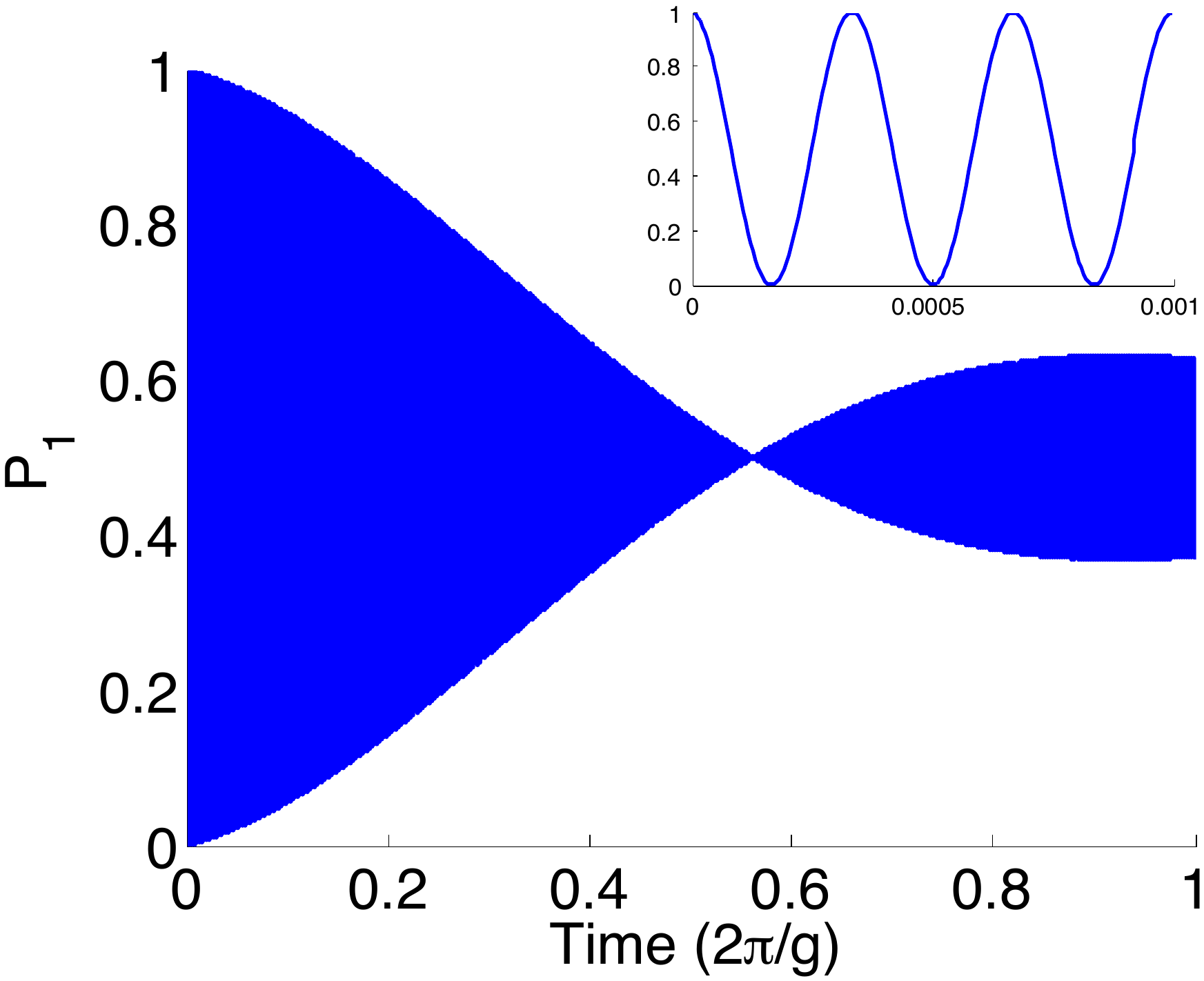}
\caption{\label{p1lab} (Color online) Probability of finding the flux qubit in the excited state as a function of time (in units of $2\pi/g$), for a spin initial state $\ket1$. Time evolution is computed in the laboratory frame. The plot is at resonance condition, decoherence is $T_1=20$ $\mu$s, $T_\nu=15$ $\mu$s. Inset: zoom on the fast Rabi oscillations for very short times.}
\end{figure}

The flux qubit undergoes two independent non-dissipative evolutions: the first is due to the microwave drive, at frequency $\Omega$, the second to the coupling with the NV center at frequency $g/2$. The latter is the one we are interested in while the former is included in the free Hamiltonian $\hat H_0$. We derive the master equation based on these ingredients in the so-called Born-Markov approximation~\cite{carmichael}. The Lindblad equation should model the effect of damping and dephasing in the basis of the free Hamiltonian eigenstates. Since the free Hamiltonian of the flux qubit is proportional to $\hat \sigma_x$, the Lindblad operators will be composed of raising/lowering operators in the Pauli-$\hat \sigma_x$ basis. Specifically, we get the following dynamical equation for the system density matrix in the laboratory frame
\begin{align}\label{dynequa}
\frac{\mathrm d\hat\rho}{\mathrm dt}&=-i[\hat H_0+\hat H_{\mathrm{eff}},\hat\rho]\notag\\
+\Gamma_x&\mathcal L[\hat \sigma_x^{fq}]\hat\rho+\Gamma_-\mathcal L[\hat \sigma_{-,x}^{fq}]\hat\rho+\Gamma_+\mathcal L[\hat \sigma_{+,x}^{fq}]\hat\rho
\end{align}
where $\mathcal L$ is the Lindblad superoperator $\mathcal L[\hat a]\hat\rho=\hat a\hat\rho \hat a^\dagger-\frac12(\hat a^\dagger \hat a\hat\rho+\hat\rho \hat a^\dagger \hat a)$. Notice that in contrast with the standard approach the microwave drive rotates the decoherence basis: the ground state is a thermal distribution in the $\{\ket-,\ket+\}$ basis. Even though the environment was assumed to be at zero temperature, the driving field leads to an effective heating of the bath. Moreover, the $\Gamma_i$'s decay rates are related to environmental spectral properties and thus to experimentally accessible quantities~\cite{ithier}. Let us define $T_1$ as the energy relaxation time and $T_\nu$ the Rabi oscillations decoherence time which can be measured through independent experiments. Then we have:
\begin{equation}
\Gamma_x=\frac14T_1^{-1},\ \Gamma_-=\frac14T_1^{-1},\ \Gamma_+=\frac14T_1^{-1}+T_\nu^{-1}
\end{equation}
One should notice the asymmetry between $\Gamma_+$ and $\Gamma_-$ that will have peculiar consequences later on. With the dynamical equation~(\ref{dynequa}) we are able to numerically evaluate the time evolution of the system, including in our simulations realistic values for $T_1$ and $T_\nu$, experimentally determined {\it e.g.} in~\cite{stern14}. There we have $T_1=10-20$ $\mu$s and $T_\nu=10-15$ $\mu$s. In Fig.~\ref{p1lab} we show the effect of decoherence on the reduced flux qubit state obtained after tracing over the spin state. We plot the probability of finding the flux qubit in the excited state, $P_1(t)={\rm Tr_{spin}}(\bra{1_{fq}}\hat \rho(t)\ket{1_{fq}})$, where ${\rm Tr_{spin}}$ denotes the partial trace over the spin's degree of freedom.

We will now show how this formalism can be used in fundamental quantum information protocols.

\section{Applications in quantum information protocols}\label{Prop}

We will see in the following how a proper choice of the parameters $\alpha$ and $\phi$ can lead to a number of applications relying on the coupling between the two-level systems, such as manipulation and measurement of the spin's state.

\subsection{Detecting spin-qubit coupling}\label{detecting}

The first step in view of exploring the flux qubit-single spin coupling is to detect the coupling itself. Here we study a protocol enabling this in presence of decoherence, which introduces noise and loss of information about the flux qubit's state. 

\begin{figure}[h]
\includegraphics[scale=.24]{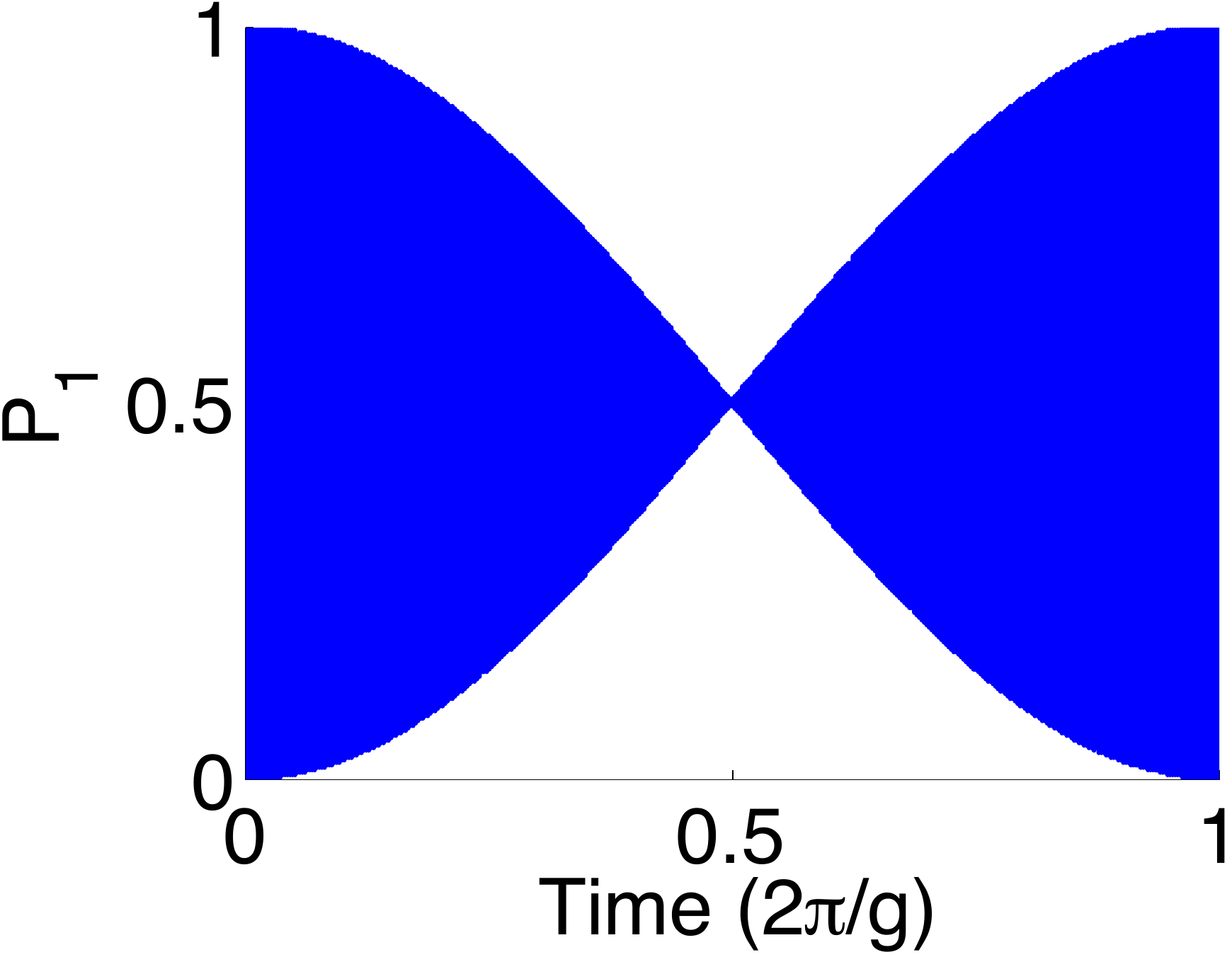}
\includegraphics[scale=.24]{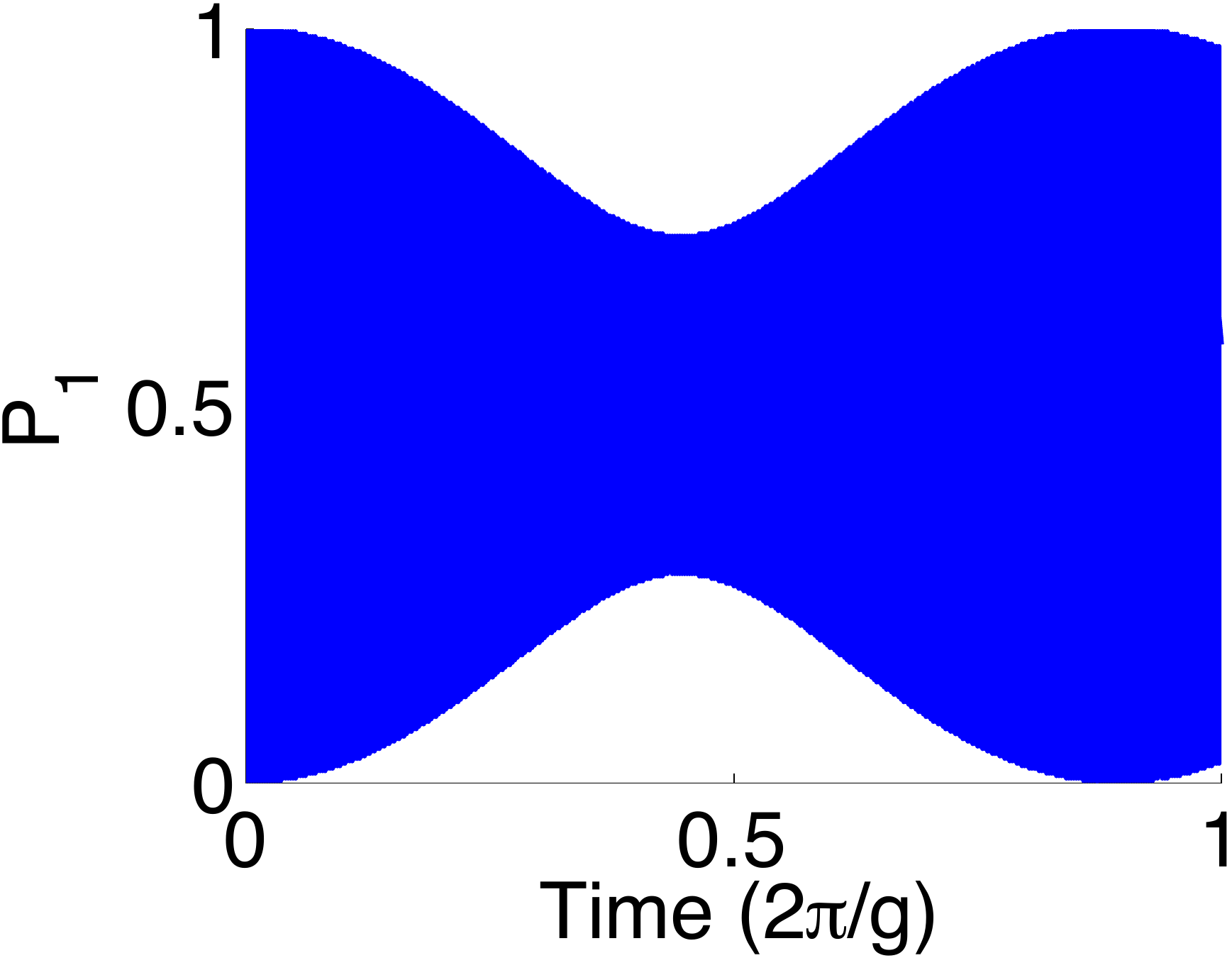}
\includegraphics[scale=.24]{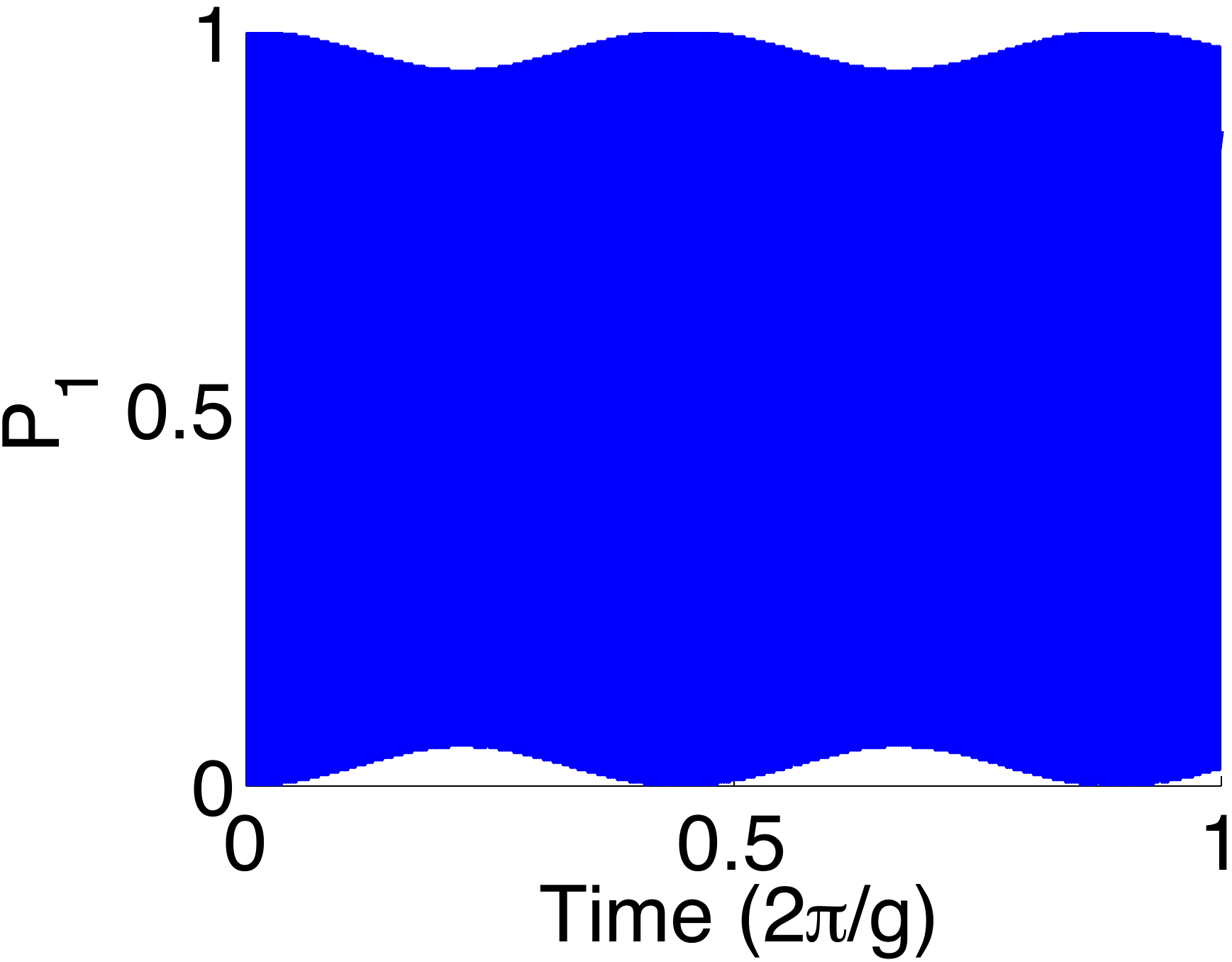}
\includegraphics[scale=.24]{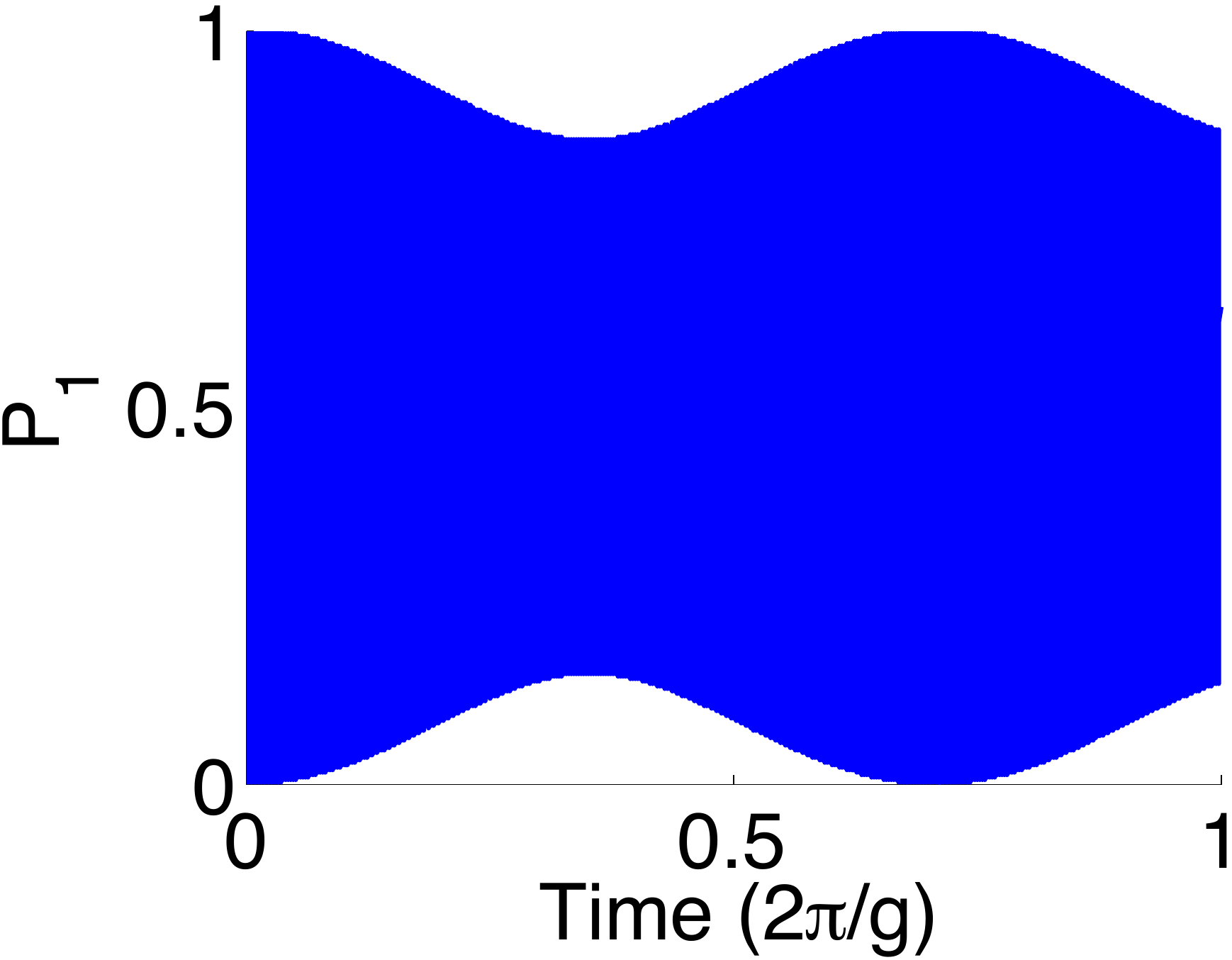}
\caption{(Color online) Probability of finding the reduced flux qubit density matrix in the excited state $\ket1$ as a function of time (in units of $2\pi/g$). Decoherence here is not taken into account. From top left and increasing clockwise, values of the detuning are: $\Omega+\delta=0,g/2,g\,{\rm and }\,2g$. \label{OscNoDec}} 
\end{figure}

We first provide the exact calculation neglecting dissipation and restricting to zero detuning, i.e. $\Omega=-\delta$, focusing on the interaction picture. We define $P_1^i(t)={\rm Tr_{spin}}(\bra{1_{fq}}\hat \rho^i(t)\ket{1_{fq}})$, where $\rho^i(t)$ is the density matrix of the whole system computed in the frame rotating with $\hat H_0$. Based on Eq.~\ref{time} we have: 
 \begin{eqnarray} \label{reduced3}
P_1^i(t)= \frac{1}{2}\left(1+\sin{2\alpha}\cos\phi\cos{\frac{gt}{2}}\right)
 \end{eqnarray}
We recall that the main goal of the present protocol is to detect the existence of the coupling. In this regard, the key point of the dynamics of the population $P_1^i(t)$ is that it does not depend on the NV center state, which is {\it a priori} unknown, but rather on the flux qubit state which is controllable. This feature will not be preserved when including decoherence to the dynamics. Hence we can set  $\alpha=\pi/4$ ({\it i.e.} starting with the flux qubit in the excited state $\ket1$) to get the highest visibility. In that case $P_1^i(t)$ reads:
\begin{equation}\label{p1}
P_1^i(t)= \frac{1}{2}\left(1+\cos{\frac{gt}{2}} \right)
\end{equation}
This equation is valid provided the flux qubit and the spin are in perfect resonance $\Omega=-\delta$.  Furthermore, the resonance condition ensures that $P_1^i(t)$ corresponds to the enveloppe of the fast oscillations when moving back to the laboratory frame (see Fig~\ref{OscNoDec}, top left). Eq.~(\ref{p1}) corresponds to an oscillation with unit visibility. The fast oscillations at frequency $\Omega$ come from the classical driving field which induces the free Hamiltonian of Eq.~\ref{lab}. They are therefore of no interest regarding the coupling with the spin.

In practice, exact resonance must be determined experimentally. In this respect, driving the flux qubit off-resonance reduces the visibility of the coupling oscillations, as can be seen in Fig.~\ref{OscNoDec}. So one practical way to find out the resonance is by modifying $\Omega$ and searching for high visibility oscillations. It is possible to gradually vary $\Omega$ until the slowly oscillating terms reach a population inversion dynamics -- practically cross the $P_1=1/2$ horizontal line, which uniquely characterizes the resonance condition. Fig.~\ref{OscNoDec} shows the time evolution of $P_1(t)$, defined in section~\ref{Decoherence}, for different values of the detuning. It is clearly possible to identify when the resonance condition is matched. Moreover, the amplitude of the oscillations characterizing the coupling are very sensitive to the detuning -- typically for values as low as $g/2$ -- which guarantees the precision of this protocol.

\begin{figure}[h]
\includegraphics[scale=.24]{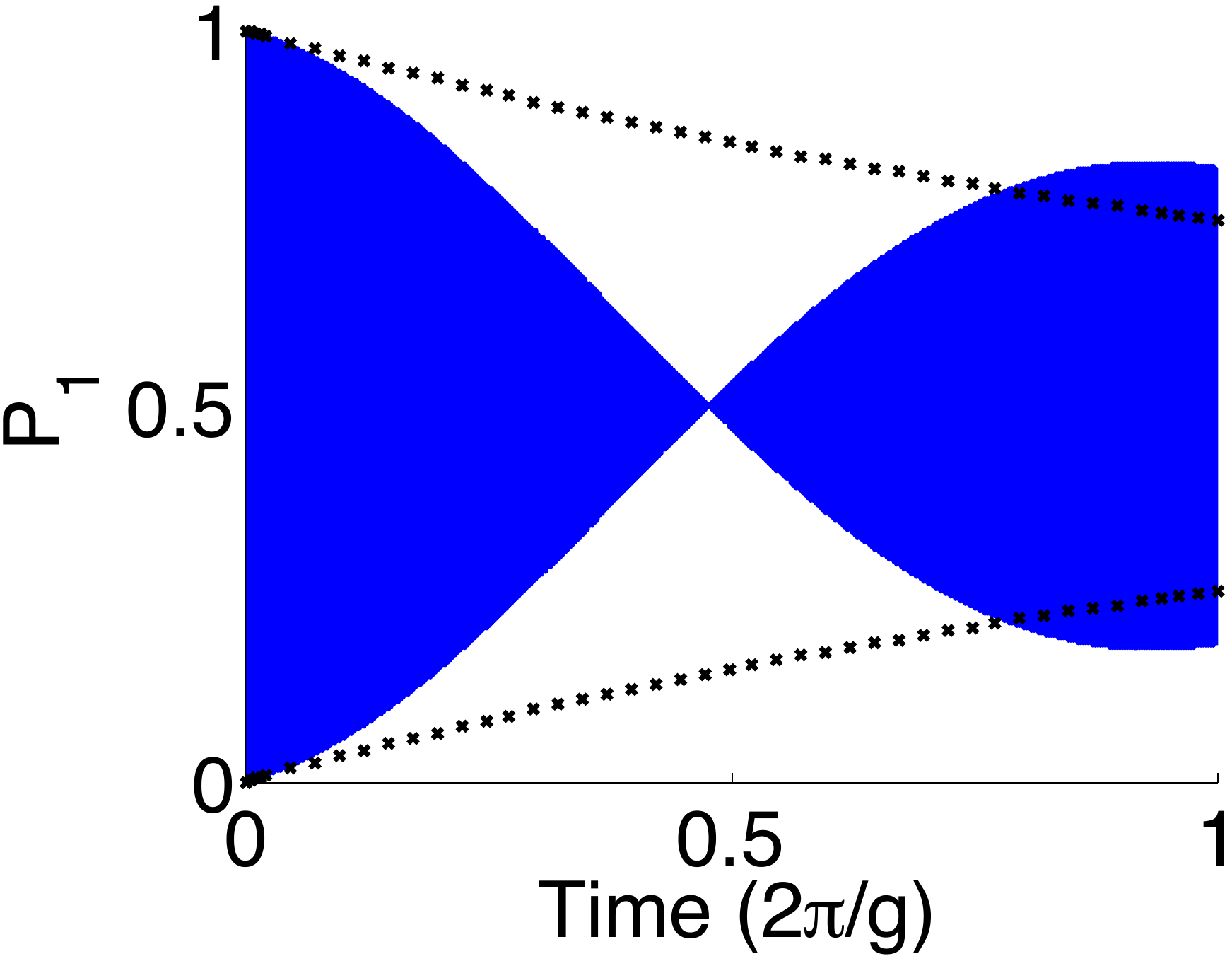}
\includegraphics[scale=.24]{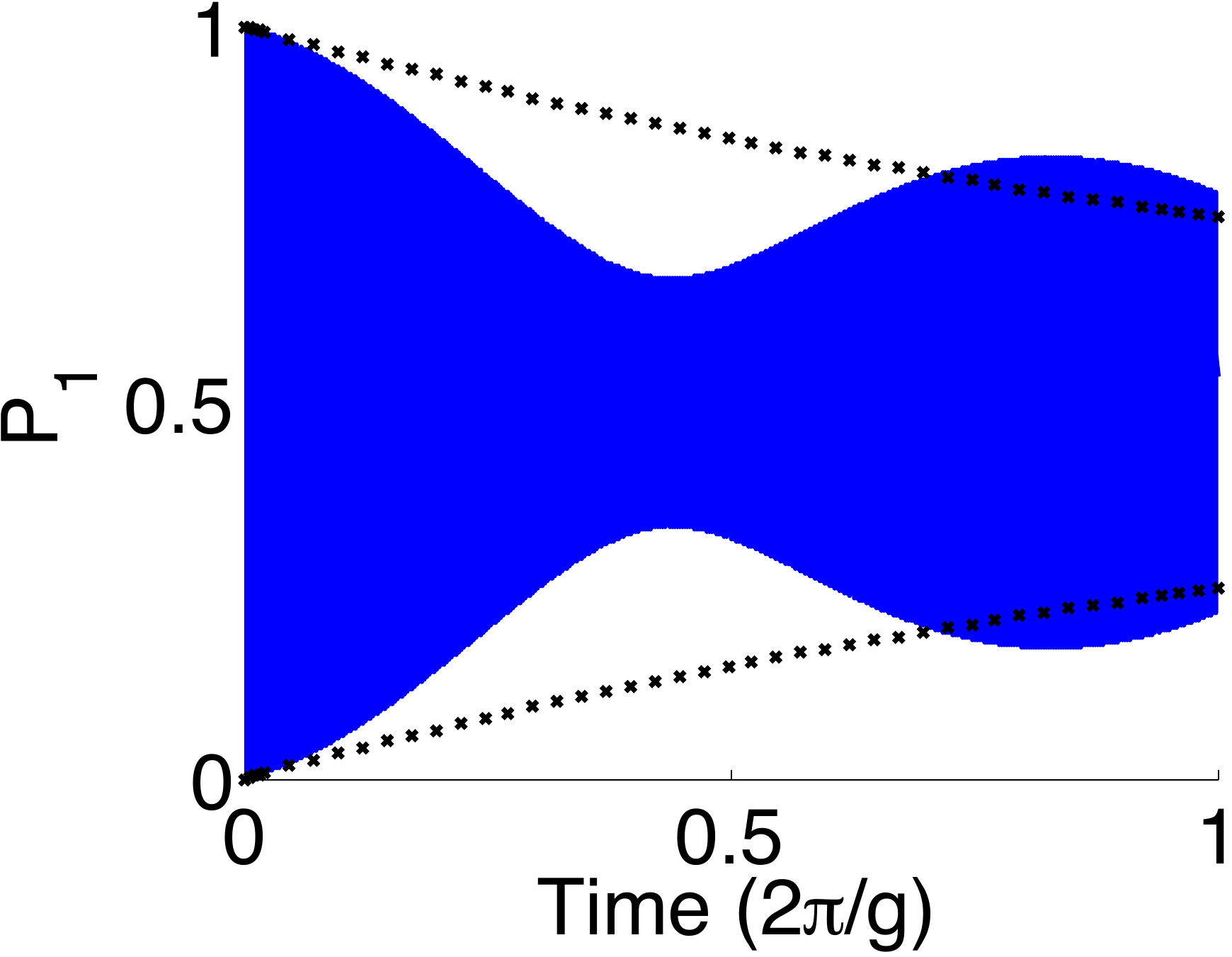}
\includegraphics[scale=.24]{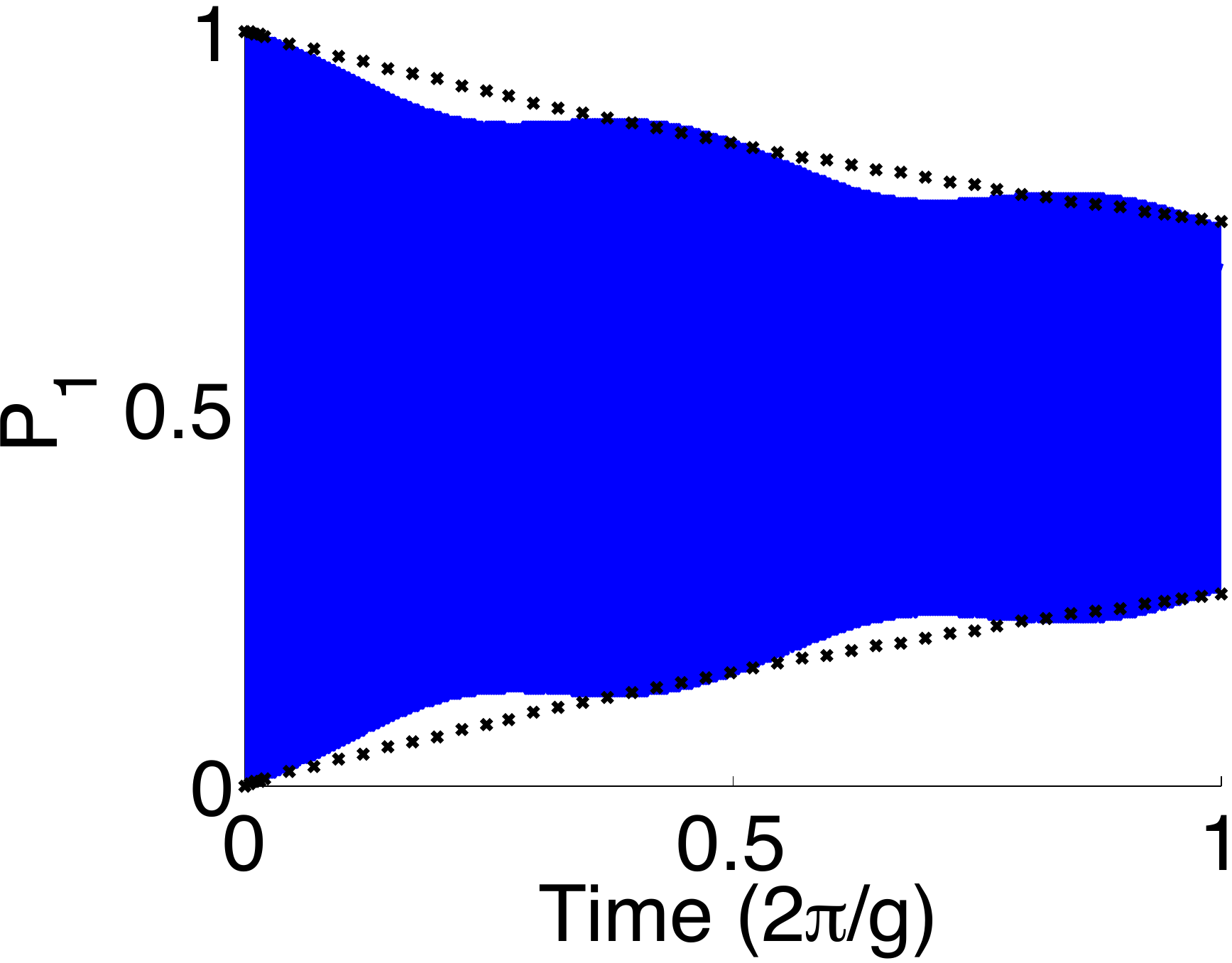}
\includegraphics[scale=.24]{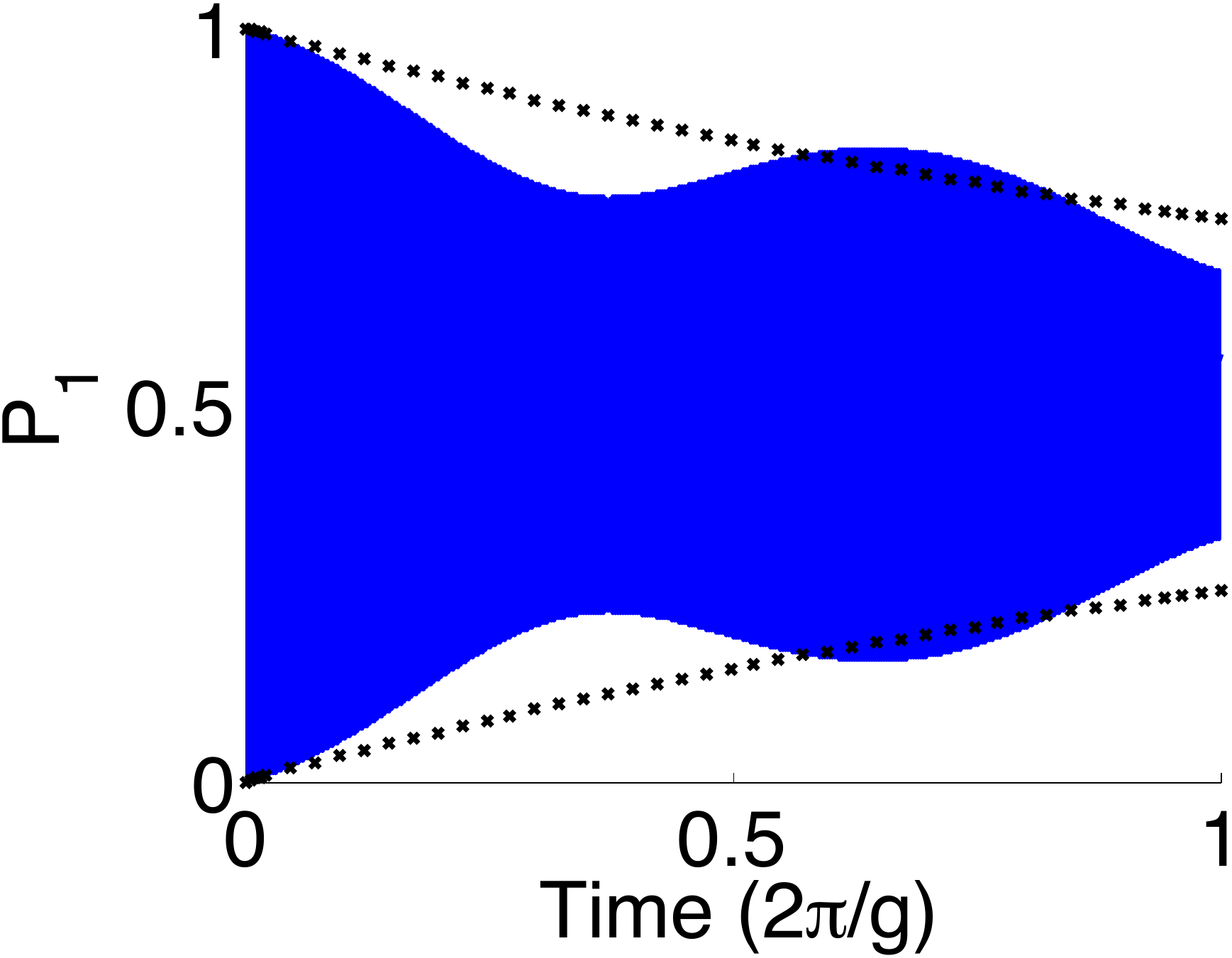}
\caption{(Color online) Probability of finding the reduced flux qubit density matrix in the excited state $\ket1$ as a function of time (in units of $2\pi/g$). Decoherence is modeled through $T_1=20$ $\mu$s, $T_\nu=15$ $\mu$s. Black crosses show the enveloppe of the fast oscillations in the absence of coupling $g=0$. From top left, and increasing clockwise, values of the detuning are : $\Omega+\delta=0,g/2,g\ {\rm and }\ 2g$. \label{OscDec}} 
\end{figure}

However, decoherence may significantly damp the oscillations and destroy the expected signal related to the population measurement of state $\ket{1}$. Furthermore, as will be shown later, the dynamics gets to depend on the spin's initial state. In order to characterize the effects of decoherence we studied the behavior of $P_1(t)$ in the presence of decoherence for different values of the system coherence times and detuning (see Fig.~\ref{OscDec}). We set the initial state of the spin to the ground state $\ket0$ and based the numerical analysis on the Lindblad equation derived in Eq.~\ref{dynequa}. It is clear from Fig.~\ref{OscDec} that even when including decoherence, the coupling of the flux qubit with the spin can be identified through the measurement of oscillations in the excitation probability of the flux qubit. The sensitivity to the detuning is sustained, which enables searching for the resonance condition by gradually varying $\Omega$.

\begin{figure}[h]
\includegraphics[scale=.24]{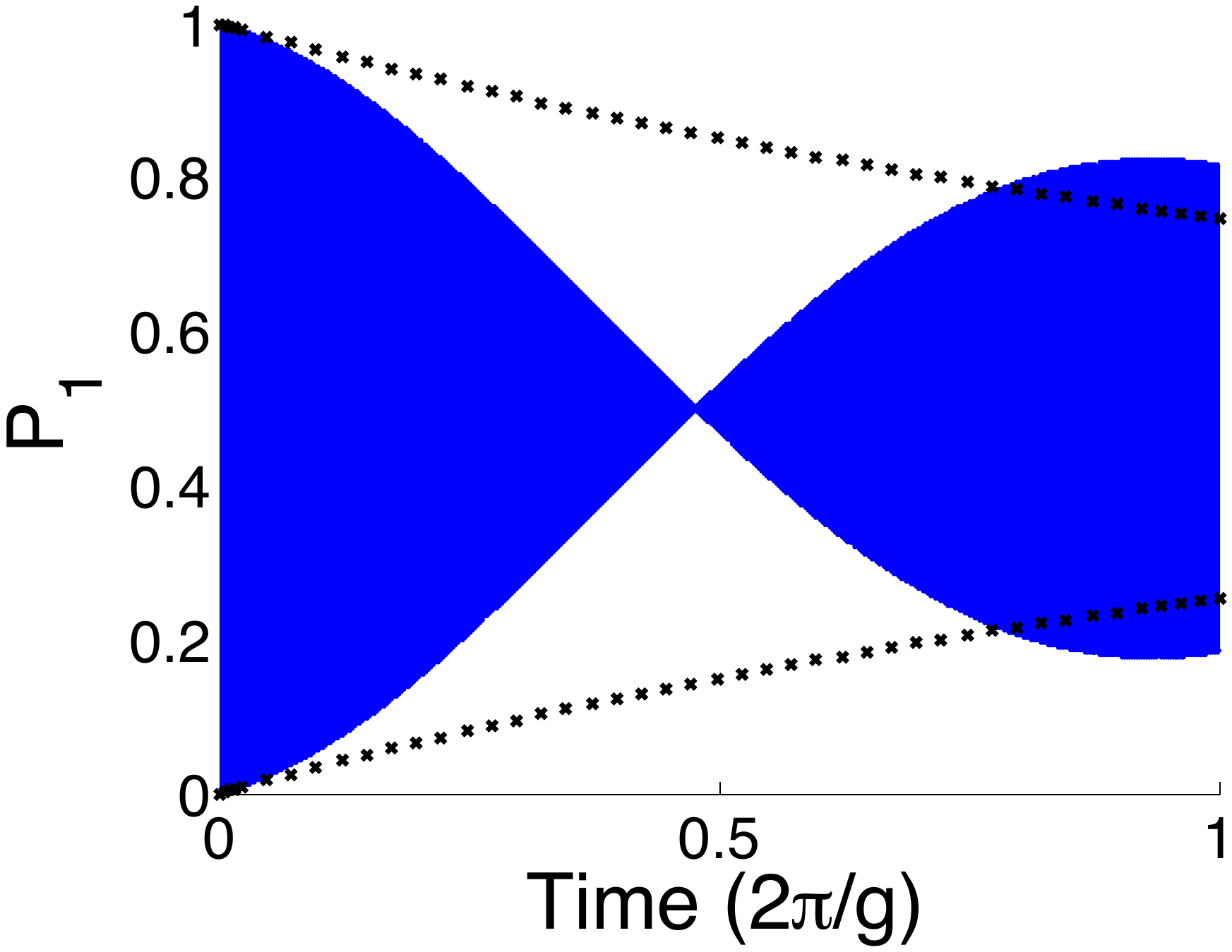}
\includegraphics[scale=.24]{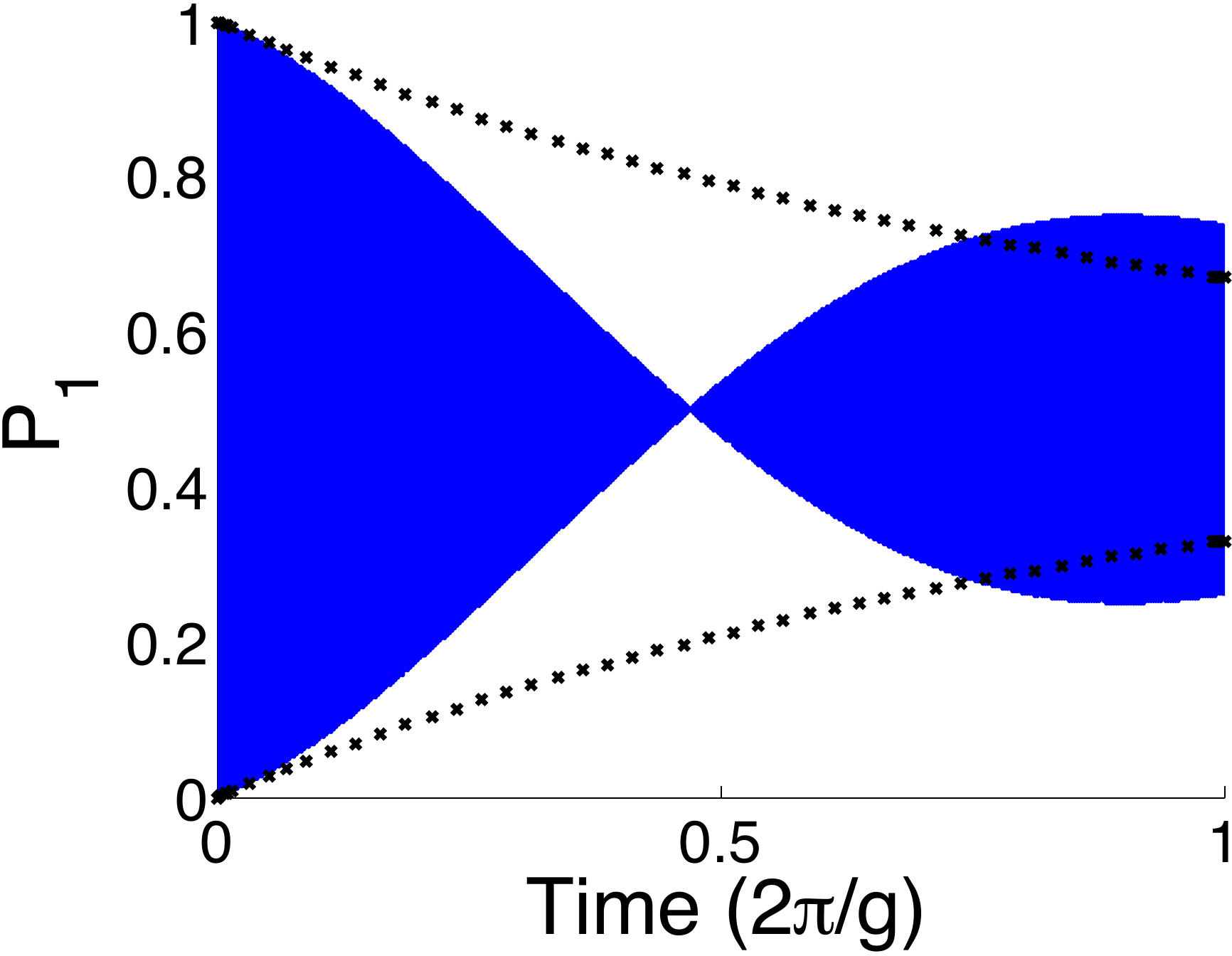}
\includegraphics[scale=.24]{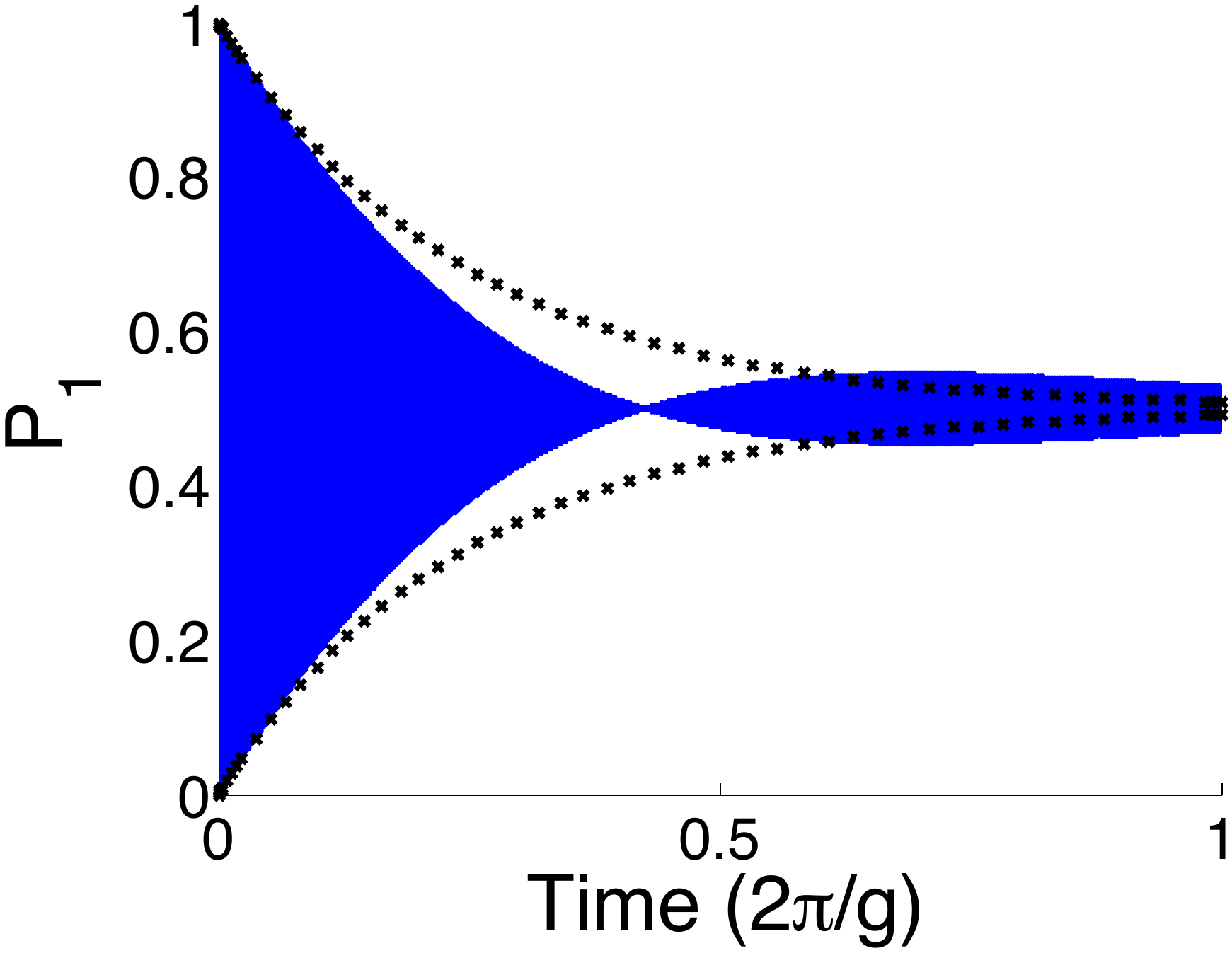}
\includegraphics[scale=.24]{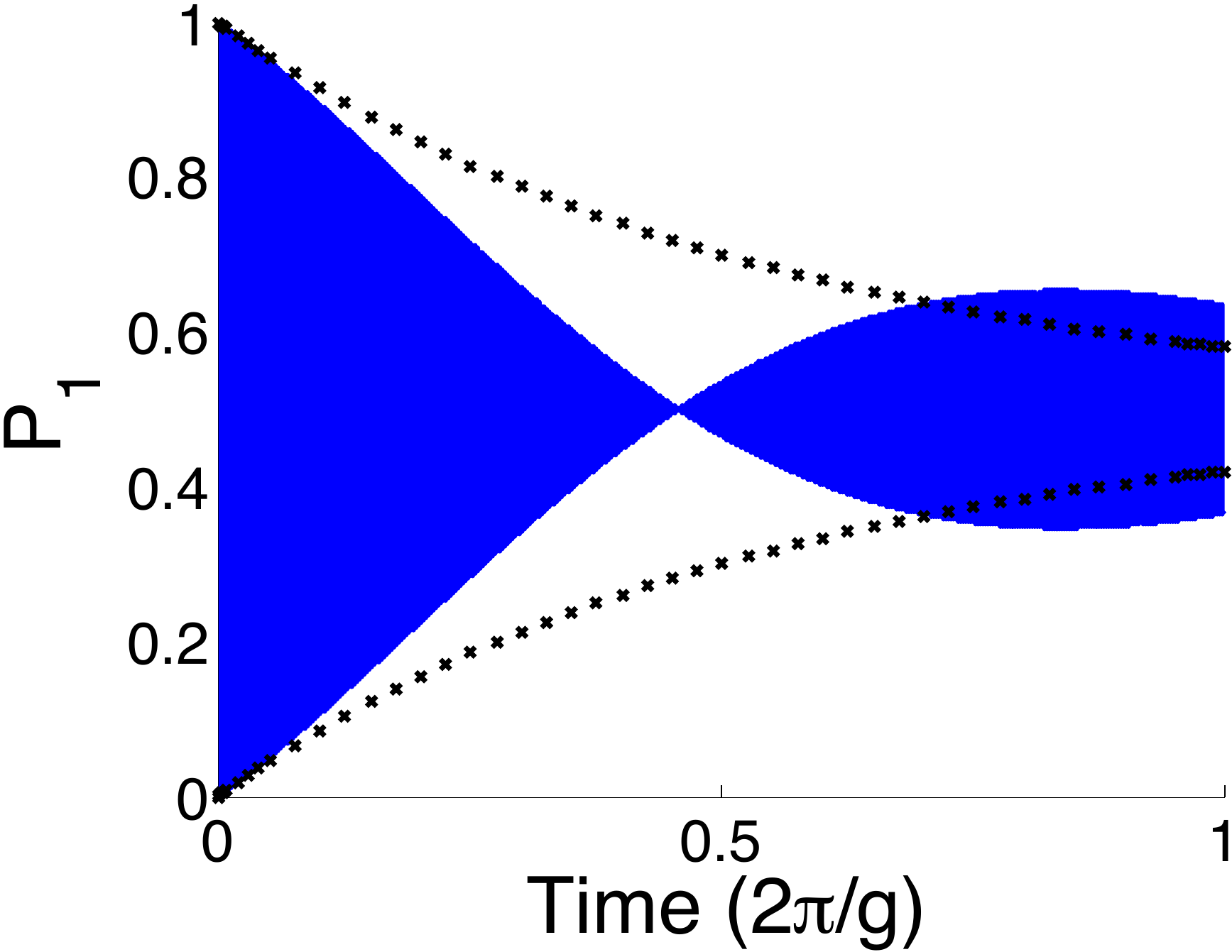}
\caption{(Color online) Probability of finding the reduced flux qubit density matrix in the excited state $\ket1$ as a function of time (in units of $2\pi/g$). $T_\nu=15$ $\mu$s while $T_1$ is 20, 10, 5 and 2 $\mu$s (from top left, decreasing clockwise). Black crosses show the enveloppe of the fast oscillations in the absence of coupling $g=0$.\label{figT1}} 
\end{figure}

One may ask how low the coherence times can be in order to see the slow oscillations due to the coupling with the spin. The coupling should be strong enough with respect to decoherence to allow for the experimental observation of at least one oscillation. In Fig.~\ref{figT1}, we show the oscillations at resonance condition for decreasing values of $T_1$, while $T_\nu$ is set to 15 $\mu$s. The initial state is $\ket{0_s}\ket{1_{fq}}$. Surprisingly enough, the figures reveal that the coupling can be identified for rather low values of $T_1$ such as 2 $\mu$s, that is $gT_1\simeq0.2$.

\begin{figure}[h]
\includegraphics[scale=.24]{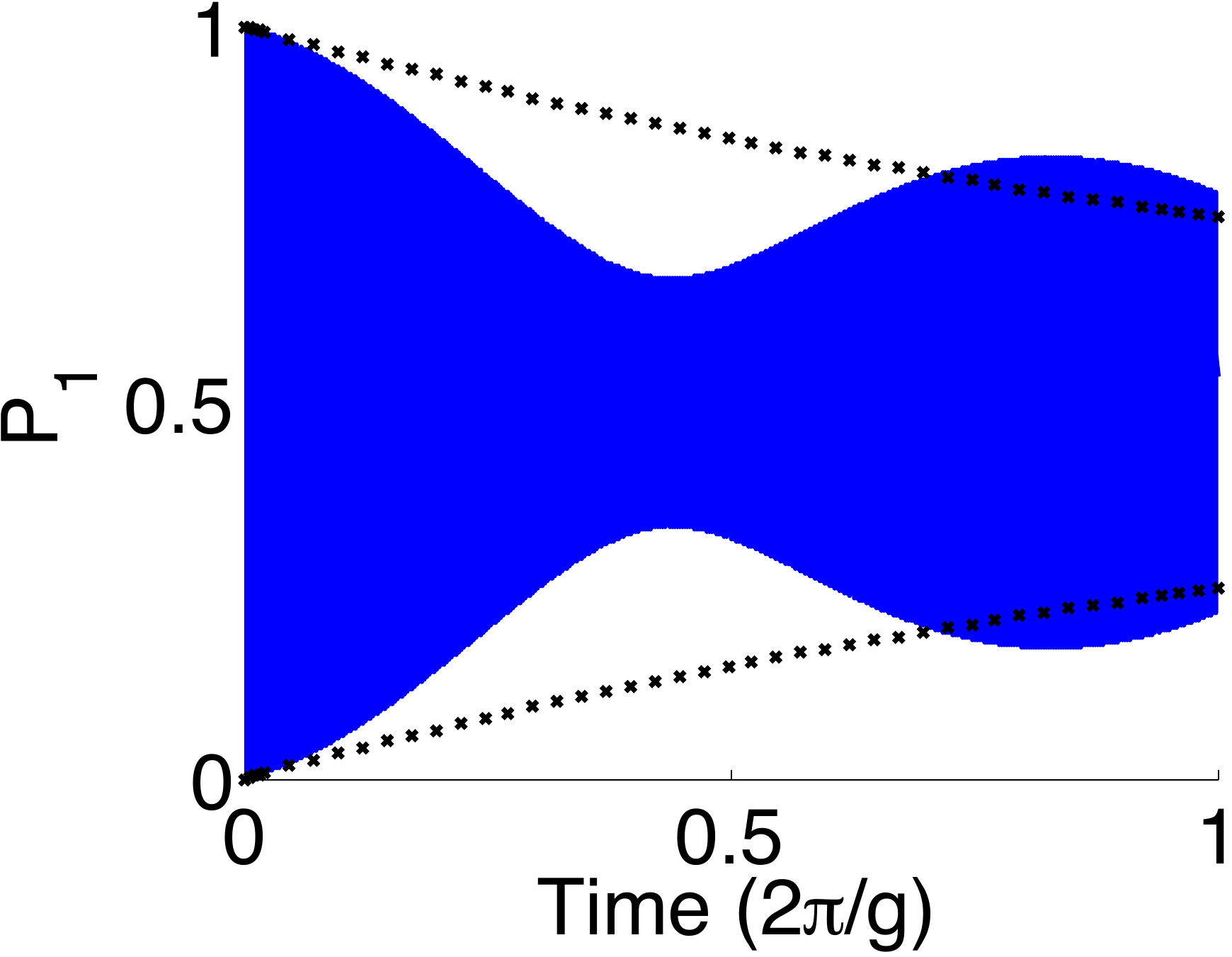}
\includegraphics[scale=.24]{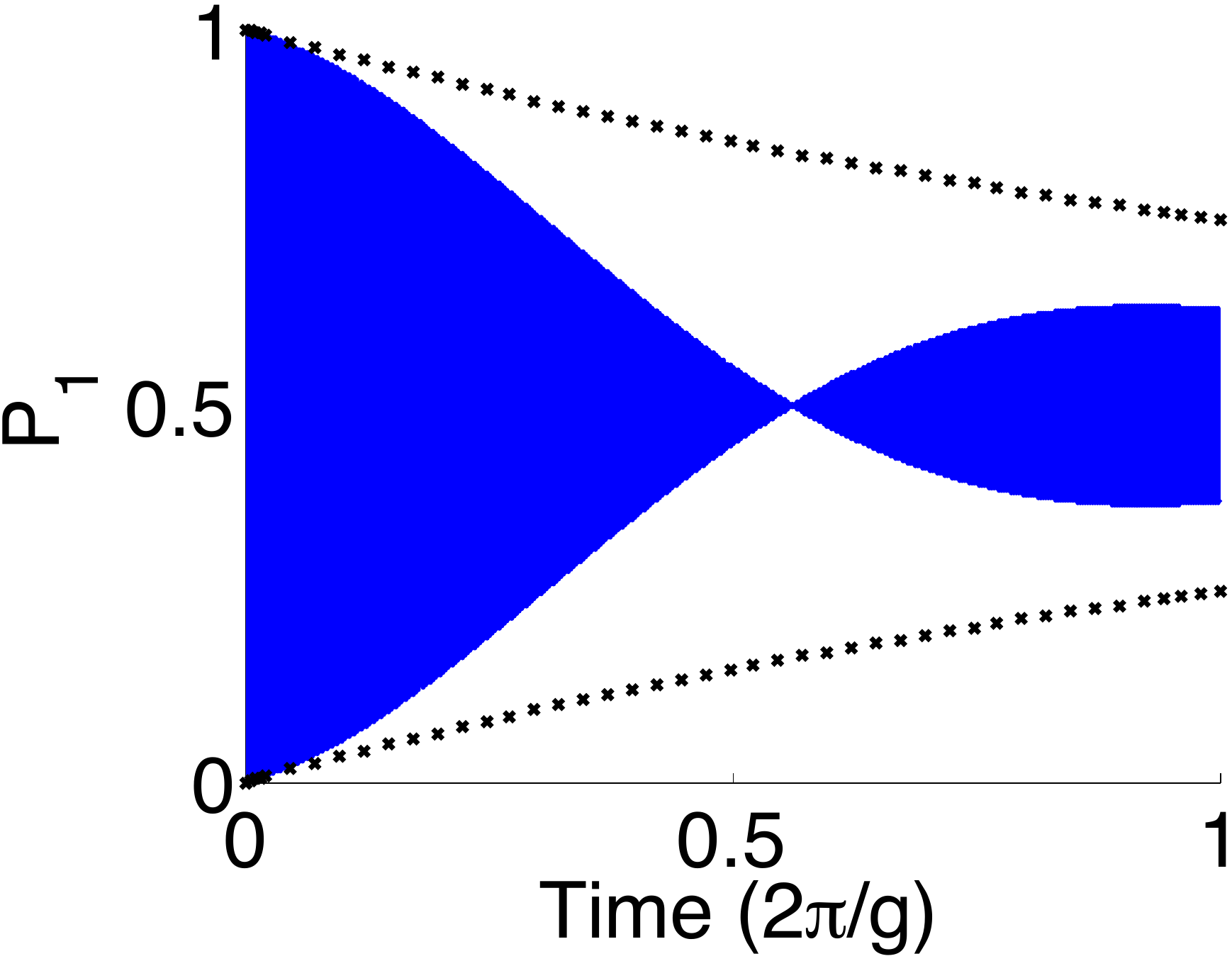}
\caption{(Color online) Probability of finding the reduced flux qubit density matrix in the excited state $\ket1$ as a function of time (in units of $2\pi/g$). Decoherence is modeled through $T_1=20$ $\mu$s, $T_\nu=15$ $\mu$s. Black crosses show the enveloppe of the fast oscillations in the absence of coupling $g=0$. Left: spin initial state is $\ket0$. Right: spin initial state is $\ket1$. \label{initialState}} 
\end{figure}

An interesting feature of the coupling also appears when dissipation is considered: depending on the spin initial state, it can either protect the flux qubit from the effects of decoherence or enhance them. This effect is illustrated in Fig.~\ref{initialState} where two different initial states of the spin are considered, leading to different damping behaviors. This is a consequence of the specific form of the master equation Eq.~(\ref{dynequa}), which exhibits an asymmetry between the dissipation terms corresponding to $\sigma_{+,x}$ and $\sigma_{-,x}$. We could indeed verify that the dependence on the spin's initial state vanishes when artificially imposing $\Gamma_+=\Gamma_-$.

\subsection{Spin state initialization }\label{stateini}

A first application of the coupling described in (\ref{Heff}), that is also the first of the DiVincenzo \cite{DiVincenzo} criteria to define a qubit, is the spin state initialization. By setting $\alpha=0$ in (\ref{init}), we have  that the initial state of the hybrid system is given by
\begin{equation}\label{protein}
\ket{\psi(0)}=(\cos{\theta}\ket{0}+e^{i\varphi}\sin\theta\ket{1})\ket{+}.
\end{equation}
From the time evolution induced by the coupling between the spin and the flux qubit, after $t=\pi/g$, this state will be set to state $\ket{0}$. From this point on, the NV center can either be manipulated and set to an arbitrary state, or can be used as a quantum memory, as developed in the next subsection.

We now analyze the impact of decoherence on the initialization protocol. The results are shown in the left column of Table~\ref{tabini}. Even though we consider the NV center itself to be perfectly isolated from the environment, the purity is reduced by the interaction with the flux qubit and leads to the preparation of non-pure state instead of the target one $\ket{0}$. In Table~\ref{tabini} we call fidelity ${\rm Tr_{flux}}(\bra{0_s}\hat\rho(\pi/g)\ket{0_s})$, where ${\rm Tr_{flux}}$ denotes the partial trace over the flux qubit's degree of freedom. In the presence of high decoherence rates, one way to improve the state initialization is by repeating the protocol. After a spin-flux qubit interaction time of $t=\pi/g$, one measures the flux qubit state disregarding the results of the measurement. Mathematically, this measurement projects the flux qubit and destroys entanglement with the spin, which corresponds to tracing over the measurement eigenbasis. Then one may again prepare the flux qubit in $\ket{+}$. We then let the spin and the flux qubit interact for the same duration. The repetition of the protocol improves the state initialization procedure, until it reaches an asymptotic value independent of the initial spin state -- right column of Table~\ref{tabini}.

The values of the fidelity between the final reduced density matrix of the spin and the projector on its ground state -- see Table~\ref{tabini} -- were numerically calculated based on the dynamical evolution equation~(\ref{dynequa}). In order to compute those values, we repeated the protocol described in the former paragraph and then averaged the fidelity we obtained over 500 Haar-random initial spin states.

\begin{table}[h]
\begin{tabular}{|c|c|c|}
 \hline
  Typical times  & Average fidelity & Average fidelity  \\
  $(T_1,T_\nu)$ (in $\mu$s) & of the protocol & after 5 iterations\\
  \hline
 $(10,10)$ & 0.92 & 0.95  \\
  \hline
  $(20,15)$ & 0.96 & 0.97  \\
  \hline
\end{tabular}
\caption{\label{tabini} Average fidelities for the initialization protocol. Details on how they were computed can be found in the main text.}
\end{table}

\subsection{Quantum memory and spin state manipulation}

We have shown in Section~\ref{stateini} how to initially prepare the spin in  state $\ket{0}$. From this initialized state, we show now how the spin can be used as a quantum memory, encoding in a long lifetime quantum system the state of the flux qubit. We will also see how to adapt this strategy  to manipulate the spin's state, realizing arbitrary single-qubit rotations. 

\begin{figure}[h]
(a)\includegraphics[scale=0.42]{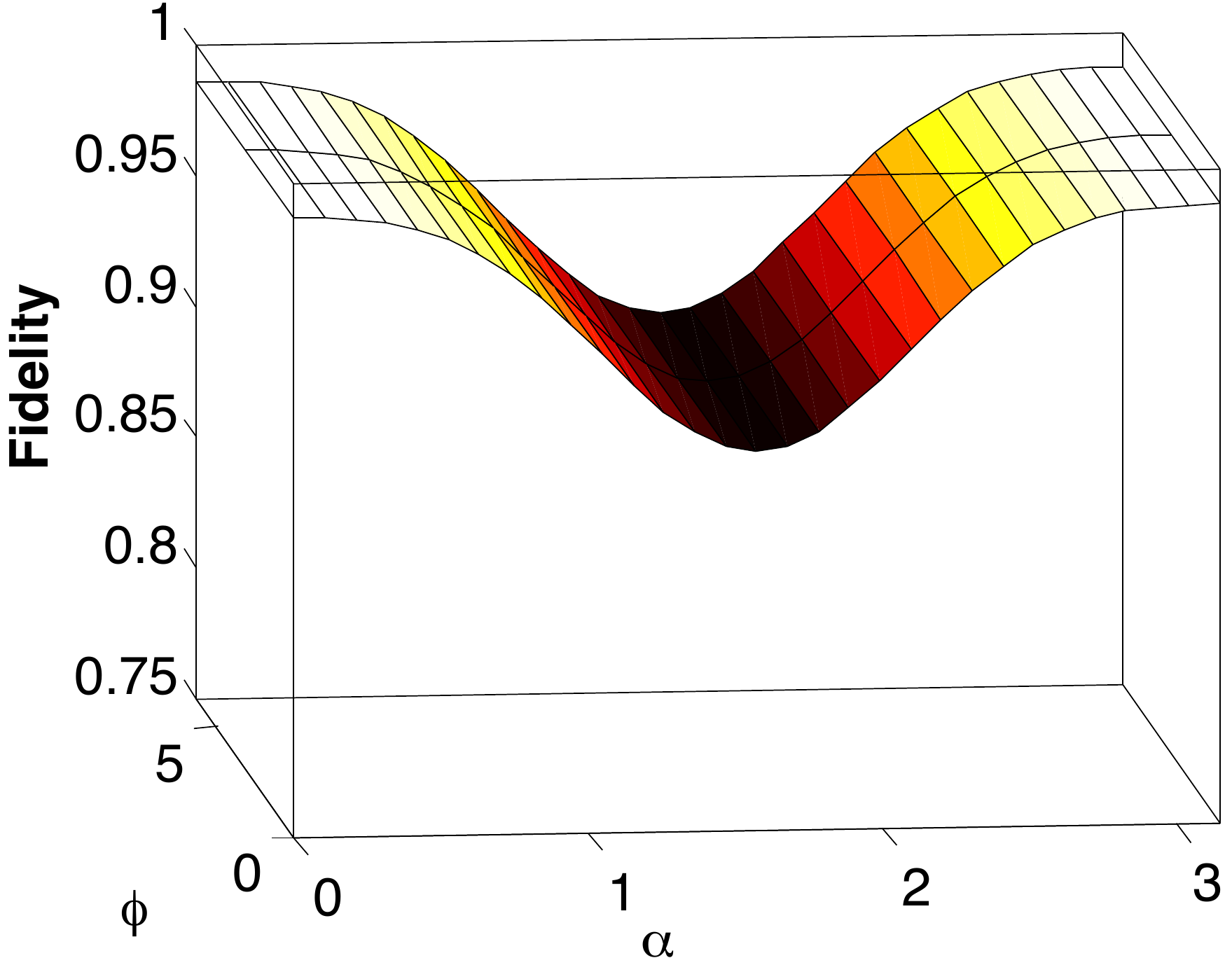}
(b)\includegraphics[scale=0.42]{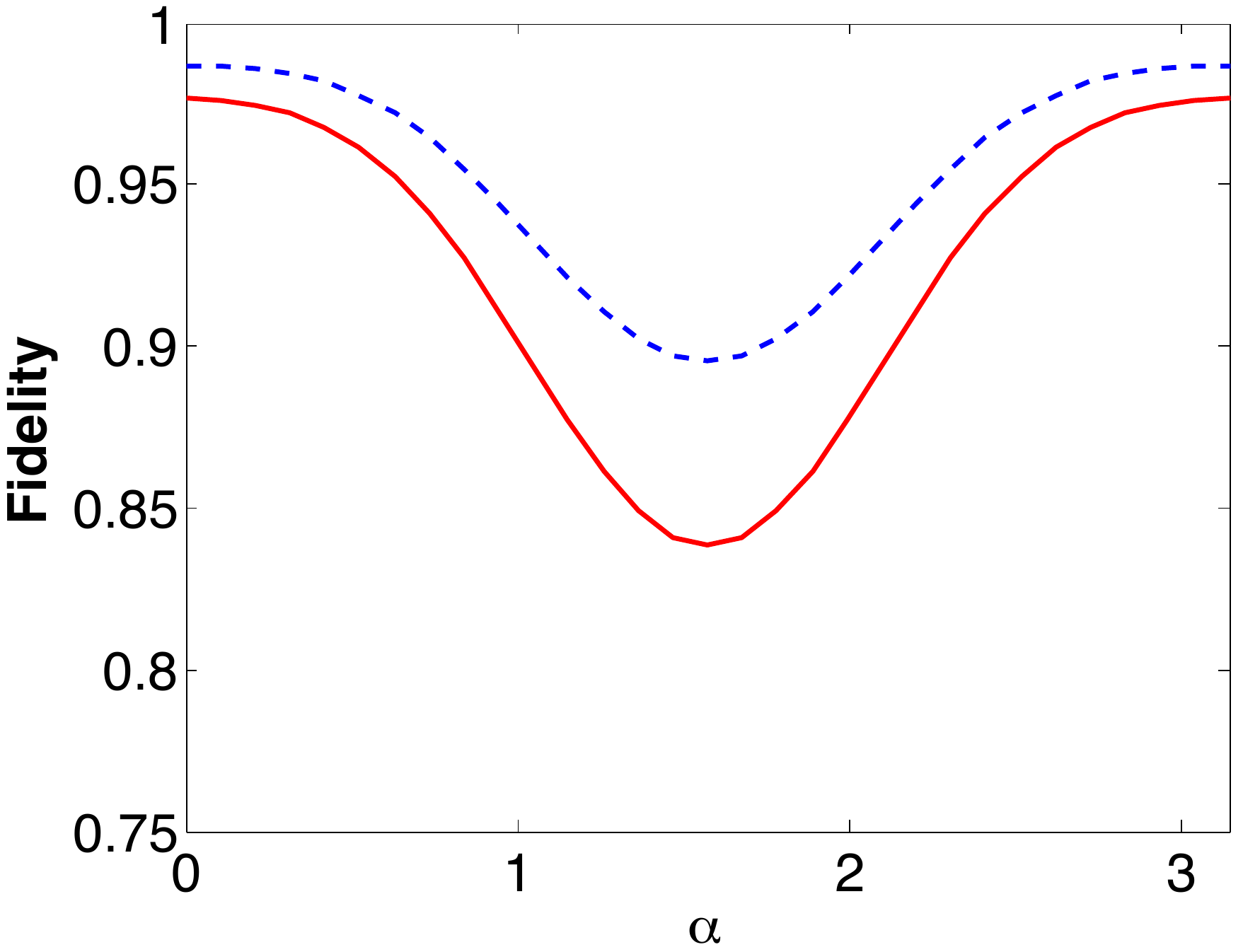}
\caption{\label{memory} (Color online) Fidelity of the NV center reduced density matrix after the protocol with the initial flux qubit state (a) as a function of $\alpha$ and $\phi$ (b) as a function of $\alpha$ only. (a) shows that the fidelity of the protocol is actually $\phi$-independent. Values of decoherence for the two plots in (b) are the same as in Table \ref{tabini}.}
\end{figure}

We start with the quantum memory protocol. If the initial spin state is $\ket{0}$, we have that $\theta=0$ in (\ref{init}) and the flux qubit is prepared in an arbitrary state that should be perfectly transferred to the spin. 
\beq \label{initmemo}
\ket{\psi(0)}=\ket{0}(\cos{\alpha}\ket{+}+{\rm e}^{i\phi}\sin{\alpha}\ket{-}). 
\eeq
Using (\ref{time}), we see that after a time $t=\pi/g$ with the initial condition (\ref{initmemo}), the SC quantum state will be completely transferred to the spin state. Thus, the latter  can play the role of a quantum memory, since it has a longer coherence time than the flux qubit. Figure \ref{memory} shows the fidelity of the final spin state after this protocol with respect to the initial flux qubit state when decoherence is also considered. The computed fidelity does not depend on the phase $\phi$, because the decoherence process itself is $\phi$-independent. The maximum fidelity is obtained for $\alpha=0$, that is for an initial state $\ket0\ket+$ which is an eigenstate of the coupling Hamiltonian of eigenvalue 0. The minimum is for $\alpha=\pi/2$ which implies a complete excitation transfer and is therefore the most likely to be impacted by decoherence.

The strategy above enables the spin state manipulation as well. Suppose that one wants to apply a given rotation to an arbitrary spin state, that can be expressed as in (\ref{protein}). Any rotation can be associated with the following transformation of the basis states of the NV center:
\begin{eqnarray}\label{rotation}
&&\ket{0}=\ket{\tilde 0}\rightarrow \cos{\beta}\ket{0}+e^{i\chi}\sin{\beta}\ket{1}\nonumber \\
&&\ket{1}=\ket{\tilde 1} \rightarrow \cos{\beta}\ket{1}-e^{-i\chi}\sin{\beta}\ket{0}. 
\end{eqnarray}
This transformation can be achieved by the following protocol: in a first step, the NV center state is reinitialized, as described in  \ref{stateini}. Thus, the state of the final coupled system composed by the spin and flux qubit is given by:
\begin{equation}
\ket{\psi_i}=\ket{0}(\cos{\theta}\ket{+}+e^{i\varphi}\sin{\theta}\ket{-}),
\end{equation}
{\it i.e.}, the state of the spin is  completely transferred to the flux qubit.  After state transfer, the coupling between the spin and the flux qubit can be stopped by turning off the intense dressing classical pulse. The flux qubit  can then be manipulated by another classical field as follows: 
\begin{eqnarray}\label{rotationSC}
&&\ket{+}\rightarrow \cos{\beta}\ket{+}+e^{i\chi}\sin{\beta}\ket{-}\nonumber \\
&&\ket{-}\rightarrow \cos{\beta}\ket{-}-e^{-i\chi}\sin{\beta}\ket{+}. 
\end{eqnarray}
Finally, the dressing microwave field can be turned on, coupling again the spin to the flux qubit. After $t=\pi/g$, we obtain the final state:
\begin{equation}\label{rotationfin}
\ket{\psi_r}=\cos{\theta}\ket{\tilde 0}+e^{i\varphi}\sin{\theta}\ket{\tilde 1},
\end{equation}
which corresponds exactly to the realization of an arbitrary rotation to the initial state (\ref{protein}). 

\subsection{NV center state tomography}

The previously discussed strategies can also be used for realizing the full spin state tomography. Since we have shown that it is possible to transfer the -- unknown -- state of the NV-center to the flux qubit, one can, after this operation, simply realize the full tomography of the flux qubit. 

Suppose now the initial state is such that $\alpha=0$ and $\phi=0$, so the flux qubit is in $\ket+$. Then for a pulse duration $t=\pi/g$, the state becomes, according to Eq.~(\ref{time}):
\begin{equation}
\ket{\psi(t)}=\ket{0}(\cos\theta\ket++i{\rm e}^{i\varphi}\sin\theta\ket-)
\end{equation}
The unknown state of the NV-center has been transferred to the accessible flux qubit. Full tomography of the latter yields perfect knowledge about the initial state of the NV-center.

We have also studied the fidelity of this protocol in presence of decoherence, finding that it displays the same behavior as the one  of the initialization protocol. \\

\section{Conclusions}

We have shown that a hybrid system, composed of a directly inaccessible spin and a superconducting flux qubit can be effectively coupled by driving  the flux qubit with an intense classical microwave field even in the limit where both two--level systems have  far off resonant characteristic frequencies. The coupling is created by the dressing of the  flux qubit by the classical field. Such dressing leads to the possibility of tuning the dressed eigenvalues to the frequency difference between the two quantum devices, a process that can be described by an effective Hamiltonian. The possibility of coupling such devices, that present complementary advantages with respect to quantum information processing, leads to a number of applications, discussed in the present paper. We have developed protocols to manipulate the spin state, use it as a quantum memory and realize its full tomography. In all protocols, a detailed study of the effects of decoherence in the dressed system was included, establishing limits on the expected fidelities according to decoherence rates compatible  to the state of the art. Our results serve as a roadmap to promising experiments using hybrid quantum devices.  An interesting perspective is studying how the flux qubit can serve as a data bus, intermediating the coupling between the spin and the quantum field of a resonator, both in the strong and the ultra-strong limits \cite{KikeUS, Cristiano}.

\end{document}